\documentclass[letterpaper,english,reprint, aps,notitlepage,superscriptaddress, twocolumn, longbibliography]{revtex4-2}

\usepackage{times}
\usepackage[T1]{fontenc}
\usepackage[utf8]{inputenc}
\usepackage{amsmath}
\usepackage{amssymb}
\usepackage{graphicx}
\usepackage[english]{babel}
\usepackage{upgreek}
\makeatletter
\let\ORIbbl@fixname\bbl@fixname
\def\bbl@fixname#1{%
	\@ifundefined{languagealias@\expandafter\string#1}
	{\ORIbbl@fixname#1}
	{\edef\languagename{\@nameuse{languagealias@#1}}}%
}
\newcommand{\definelanguagealias}[2]{%
	\@namedef{languagealias@#1}{#2}%
}
\makeatother

\definelanguagealias{en}{english}
\definelanguagealias{EN}{english}

\usepackage[table, dvipsnames]{xcolor}

\newcommand*\subtxt[1]{_{\textnormal{#1}}}
\DeclareRobustCommand\_{\ifmmode\expandafter\subtxt\else\textunderscore\fi}

\colorlet{mylinkcolor}{RoyalPurple}
\colorlet{mycitecolor}{RoyalPurple}
\colorlet{myurlcolor}{RoyalPurple}
\usepackage{hyperref}
\hypersetup{
	linkcolor  = mylinkcolor,
	citecolor  = mycitecolor,
	urlcolor   = myurlcolor,
	colorlinks = true,
	breaklinks = true
}
\makeatletter

\pdfpageheight\paperheight
\pdfpagewidth\paperwidth

\def\dimv#1#2{#1\,\ensuremath{\mathrm{#2}}}

\def\rohg{\varrho \subtxt{hg}}
\newcommand{\vect}[1]{\boldsymbol{\mathbf{#1}}}
\def\subfig#1{\textbf{\lowercase{#1}}}

\makeatother

\begin{abstract}

\subsection*{Abstract}
Magnetic resonance imaging is a three-dimensional imaging technique, where a gradient of the magnetic
field is used to interrogate spin resonances with spatial resolution. The application of this technique to probe the coherence of atoms with good three-dimensional resolution is a challenging application. We propose and demonstrate an optical method to probe spin resonances via a two-photon Raman transition, reconstructing the 3D-structure of an atomic ensemble’s coherence, which is itself subject to external fields. Our method relies on a single time-and-space resolved heterodyne measurement, allowing the reconstruction of a complex 3D coherence profile. Owing to the optical interface, we reach a tomographic image resolution of $14\times14\times36$ $\upmu\mathrm{m}^3$. 
The technique allows to probe any transparent medium with a resonance structure and provides a robust diagnostic tool for atom-based quantum information protocols. As such, it is a viable technique for application to magnetometry, electrometry, and imaging of electromagnetic fields.

\end{abstract}

\begin{document}

\title{Coherent optical two-photon resonance tomographic imaging in three
dimensions}

\author{Mateusz Mazelanik}
\email{m.mazelanik@cent.uw.edu.pl}

\affiliation{Centre for Quantum Optical Technologies, Centre of New Technologies,
University of Warsaw, S. Banacha 2c, 02-097 Warsaw, Poland}
\affiliation{Faculty of Physics, University of Warsaw, L. Pasteura 5, 02-093 Warsaw,
Poland}
\author{Adam Leszczyński}
\affiliation{Centre for Quantum Optical Technologies, Centre of New Technologies,
University of Warsaw, S. Banacha 2c, 02-097 Warsaw, Poland}
\affiliation{Faculty of Physics, University of Warsaw, L. Pasteura 5, 02-093 Warsaw,
Poland}
\author{Tomasz Szawełło}
\affiliation{Centre for Quantum Optical Technologies, Centre of New Technologies,
University of Warsaw, S. Banacha 2c, 02-097 Warsaw, Poland}
\affiliation{Faculty of Physics, University of Warsaw, L. Pasteura 5, 02-093 Warsaw,
Poland}
\author{Michał Parniak}
\email{m.parniak@cent.uw.edu.pl}

\affiliation{Centre for Quantum Optical Technologies, Centre of New Technologies,
University of Warsaw, S. Banacha 2c, 02-097 Warsaw, Poland}
\affiliation{Niels Bohr Institute, University of Copenhagen, Blegdamsvej 17, 2100
Copenhagen, Denmark}
\maketitle

\section*{Introduction}

It has always been the primary task of optics to deliver images.
Three-dimensional (3D) imaging is now one of
the essential tools in modern sciences, medicine, and technology. The
ability to inspect the internal structure of an object not only fulfills
the basic cognitive curiosity but also constitutes a robust and direct
diagnostic method. Tomography, which nowadays appears in a plethora of
variants, revolutionized medicine and has been widely adopted
in natural and applied sciences \cite{Shin2022,Engstroem2011,Sung2021,Xiao2013}.
The key example is the magnetic resonance imaging (MRI) \cite{Lauterbur1973, Feinberg2010} that allows volumetric inspection of biological samples \cite{Lauterbur1973, Feinberg2010} by detecting space-dependent radio frequency signals generated by precessing nuclear spins placed in a magnetic field gradient. An optical version of MRI has been proposed as a technique to dramatically increase the spatial resolution of the system \cite{Allodi2016}.
In physics, 3D imaging has been used to reveal and study microscopic
features in quantum fluids \cite{Kasai2018} and solids \cite{Karpov2017,Kardjilov2008,Kim2018},
to detect and localize single spins \cite{Rugar2004,Streed2012,Willke2019,Zopes2018}
and to visualize classical \cite{Appel2015,Boehi2010,Horsley2015,Kardjilov2008}
and quantum electromagnetic fields \cite{Lee2014}.

Here we employ a cold-atoms-based memory and demonstrate a method to optically resolve a three-dimensional
spatial distribution of coherence between two atomic spin states, in its
full complex form. We directly demonstrate and benchmark
the sensitivity of our method by phase-modulating the coherence with
a predetermined pattern. Finally, we show the ability to detect
a magnetic field structure by reconstructing its phase footprint
on the coherence. 
This achievement is relevant
to all countless protocols that process quantum information carried
in such a coherence \cite{Lvovsky2009,Saglamyurek2018,Duan2001}. Via this coherence, we are able to measure external fields influencing the atoms in three dimensions, which in our case include optical and magnetic fields  \cite{Castellucci2021,Xu2008}, but possibly could be extended to microwave or terahertz sensing via Rydberg states \cite{Jing2020, PhysRevX.10.011027}, imaging of interactions \cite{Sompet2022, Ferreira-Cao_2020}, or as an alternative technique and a three-dimensional extension to quantum gas microscopes \cite{Bakr2009, Cheuk2015}. The technique could also be mapped to solid-state resonant systems that can be optically probed, including the example of color centers in diamond \cite{Balasubramanian2008, Gruber1997}.
Furthermore, our method is a practical tool for testing optical quantum memories, providing means to verify phase homogeneity, which is particularly relevant if specific interference of emissions from all atoms is desired, such as for example in the case of superradiance. 

\begin{figure*}
\centering
\includegraphics[width=1\textwidth]{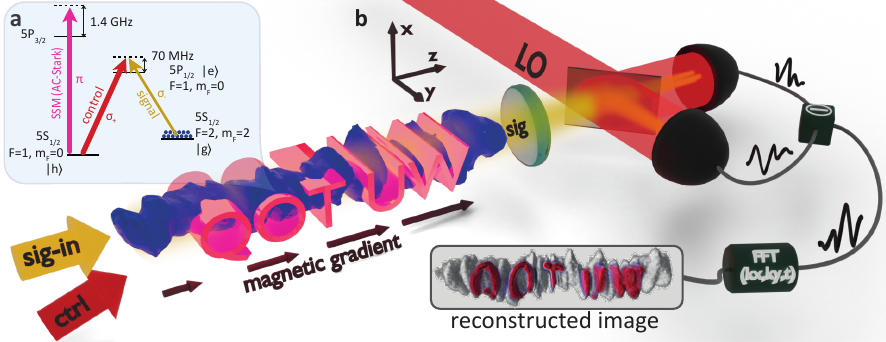}\caption{\textbf{Experimental setup for 3D phase-sensitive atomic coherence imaging.}\ \subfig{a} Light-atoms
interface used to generate, modulate and retrieve the atomic coherence
(spin wave). The right-hand circularly polarized ($\sigma_{+}$) control (red) and left-hand circularly polarized ($\sigma_{-}$) signal (yellow) fields couple a two-photon transition $|\text{h}\rangle\to|\text{g}\rangle$ between states of total angular momentum $\text{F}=2$ and $\text{F}=1$ respectively. This enables the mapping of the signal field onto atomic coherence. The off-resonant AC-Stark linearly polarized ($\pi$) laser (pink, SSM) is used to imprint phase profiles onto the created coherence.
 \subfig{b} The light emitted from the atomic cloud that stores a phase-modulated (QOT UW inscription) atomic coherence is detected in the far field of the ensemble using a spatially-resolved heterodyne detector enabled by the interference of the signal (sig) light with a local oscillator (LO).
The temporal shape of the emitted signal contains information associated
with the longitudinal coherence wavevector. The three-dimensional Fourier transform (represented as FFT block) of the collected heterodyne frames allows the reconstruction of the atomic coherence. The coherence can be created by sending to the atoms an input signal (sig-in) accompanied by a control field (ctrl) pulse.
\label{fig:Setup}}
\end{figure*}

\section*{Results}
\subsection*{Operating principle}
The key idea behind our method is to map the coherence to light in a way that preserves information about its three-dimensional structure. Many atom-light interfaces,
such as Electromagnetically Induced Transparency (EIT) \cite{Hsiao2018}, Autler-Townes splitting \cite{Saglamyurek2018}
or Raman scattering \cite{Guo2019} allow restoring the shape of the coherence in dimensions transverse to the propagation axis. The structure along the ensemble is however lost in the mapping process. The way to prevent this is to alter the atoms-to-light mapping process, so different components along the propagation direction are mapped to different frequency components of the emitted light. This is one of the key features of the Gradient Echo Memory
(GEM) \cite{Cho2016} protocol in which atomic transition frequencies are altered by a magnetic field gradient causing Zeeman shifts. For a linear gradient, during the readout, the coherence along the ensemble is mapped linearly to the spectrum of the emitted signal pulse. At the same time, the transverse components are directly mapped to the corresponding distribution of the optical field. In reciprocal coordinates the readout process couples Fourier components of atomic coherence with wavevector $(k_{x},k_{y},k_{z})$
to chunks of readout signal light described by coordinates $(k_{x},k_{y},t \subtxt{R})$,
i.e. emitted at a certain time with matching perpendicular wavevector
components. The longitudinal direction $z$ is both the direction of the propagation
of the signal, as well as the direction of the magnetic field gradient. The correspondence between the time $t \subtxt{R}$ and $k_{z}$ is determined by the amount of time a magnetic
field gradient $\beta=\partial_z B_{z}$ needs to decelerate the atoms with momenta
$\hbar k_{z}$ to rest. In the spectral and real space coordinates this correspondence translates directly to the linearly changing Zeeman shift caused by the gradient $\mu \subtxt{B} \beta z \leftrightarrow  \omega \hbar$, where $\mu \subtxt{B}$ is the Bohr magneton.  By measuring the amplitude and phase of the signal
light as a function of $(k_{x},k_{y},t \subtxt{R})$ all the information
on initial atomic coherence can be recovered. 


\subsection*{Protocol}

In order to recover the profile of the atomic coherence from the optical measurement, we need to first understand the intricacies behind the interaction. In essence, the spatial dependence will be recovered by means of a multi-dimensional Fourier transform. However, effects such as diffraction necessitate careful theoretical treatment.
To simplify the derivation of position-dependent coupling factors
and phases introduced by diffraction let us consider the following
protocol. At time $t=0$ we are given an atomic sample with an unknown
coherence $\varrho \subtxt{hg}(x,y,z)$ between two levels $|\text{g}\rangle$,
$|\text{h}\rangle$ (see Fig. \ref{fig:Setup}\subfig{a}) that we assume to be long-lived. The coherence must
be magnetically sensitive i.e. by applying a magnetic field gradient along the ensemble
we are able to introduce a position-dependent phase to the coherence.
Specifically:
\begin{equation}
\frac{\partial\varrho \subtxt{hg}}{\partial t}=\mathrm{i} \omega \subtxt{L}\varrho \subtxt{hg},
\end{equation}
where $\omega \subtxt{L}=\mu \subtxt{B}\beta(t)(z-z \subtxt{g})/\hbar$ is the time and space dependent
Larmor frequency, with $\beta(t)$ denoting the gradient along $z$
and $z \subtxt{g}$ the point in space where splitting vanishes (naturally
outside atomic ensemble due to bias field). This equation can be readily
solved to yield 
\begin{equation}
    \varrho \subtxt{hg}(\mathbf{r},t)=\exp(\mathrm{i} \phi(z,t))\varrho \subtxt{hg}(\mathbf{r},t=0),
\end{equation}
with $\phi(z,t)=(z-z \subtxt{g})\int_{0}^{t}\beta(t')dt'$ denoting spatiotemporal
GEM phase shift. We chose to split the total phase shift into a sum
of dominant linear terms and corrections $\phi(z,t)=\bar{\beta}zt+\bar{\omega} \subtxt{L}t+\mathcal{O}(z,t)$.

At the time $t \subtxt{R}$ we send a short strong control pulse (red arrow in Fig. \ref{fig:Setup}\subfig{b}) through the atomic
ensemble that converts the coherence $\varrho \subtxt{hg}$ to signal
light $\Omega \subtxt{s}$. The control pulse with Rabi frequency $\Omega \subtxt{C}$
and signal light with Rabi frequency $\Omega \subtxt{s}$ together drive
a two-photon transition from $|\text{h}\rangle$ through $|\text{e}\rangle$ to
$|\text{g}\rangle$, as depicted in Fig.\ref{fig:Setup}\subfig{a}. We assume that both fields are co-propagating along
$z$. In GEM protocol we work far from resonance ($\Delta/\Gamma>$
optical depth, with $\Delta$ being single photon detuning and $\Gamma$ being decay rate of the $|\text{e}\rangle$ state), thus we neglect the single photon absorption and dispersion
of the signal field. Under those conditions the growth of the signal field
$\Omega \subtxt{s}$ along atomic ensemble is governed by the equation:

\begin{align}
\frac{\partial}{\partial z}\Omega \subtxt{s} & =-\mathrm{i} gn(x,y,z)\Omega \subtxt{C}\varrho \subtxt{hg}+\frac{\mathrm{i}}{2k \subtxt{0}}\nabla_{\perp}^{2}\Omega \subtxt{s},\label{eq:dzOsxy}\\
g & =\frac{k \subtxt{0}}{\hbar\epsilon \subtxt{0}}\frac{d \subtxt{ge}^{2}}{2\Delta+\mathrm{i}\Gamma},
\end{align}
where $n(x,y,x)$ is the atomic density, $d \subtxt{ge}$ is the relevant transition dipole moment, and $k \subtxt{0}$ is the signal field wavevector length. For simplicity, we assume
that the change of coherence due to the two-photon transition, expressed
by the equation $\dot{\varrho} \subtxt{hg}=-\mathrm{i}\Omega \subtxt{s}\Omega \subtxt{C}^{*}/(4\Delta+2\mathrm{i}\Gamma)$,
can be neglected, and we combine the coherence term with atomic density
into a spin wave: $\mathcal{S}(x,y,z)=n(x,y,z)\varrho \subtxt{hg}(x,y,z)$.
Then the equation (\ref{eq:dzOsxy}) is integrated with two steps. First,
transverse dimensions $x$ and $y$ are Fourier transformed. This affects
only the diffraction term $\nabla_{\perp}^{2}\rightarrow-k_{\perp}^{2}$.
We obtain:

\begin{equation}
\frac{\partial}{\partial z}\Omega \subtxt{s}(k_{x},k_{y},z)=-\mathrm{i}\frac{k_{\perp}^{2}}{2k \subtxt{0}}\Omega \subtxt{s}(k_{x},k_{y},z)+g\Omega \subtxt{C}\tilde{\mathcal{S}}(k_{x},k_{y},z).
\end{equation}

Next, we integrate the equation along the atomic cloud, extending from $-L$
to $0$:

\begin{equation}
\Omega \subtxt{s}(k_{x},k_{y};t \subtxt{R})=
g\Omega \subtxt{C}\intop_{-L}^{0}dz\exp\left(\frac{\mathrm{i} zk_{\perp}^{2}}{2k \subtxt{0}}\right)\tilde{\mathcal{S}}(k_{x},k_{y},z;t \subtxt{R}),\label{eq:OsLkxyt}
\end{equation}
where the source term $\tilde{\mathcal{S}}$ is a Fourier transform of the spin wave along $x$ and $y$:

\begin{equation}
\tilde{\mathcal{S}}(k_{x},k_{y},z;t \subtxt{R})=\exp\left[\mathrm{i} \phi(z,t \subtxt{R})\right]\mathcal{\mathcal{F}}_{x,y\rightarrow k_{x}k_{y}}\left\{ \mathcal{S}(x,y,z)\right\}. 
\end{equation}

In equation (\ref{eq:OsLkxyt}) the $z$ integral can be extended
to infinity, since the source term is anyway nonzero only inside the atomic
cloud. The $\bar{\beta}zt$ term of GEM phase shift $\phi(z,t)$ enables
recasting the integral into a 3D Fourier transform:
\begin{multline}
\Omega \subtxt{s}(k_{x},k_{y},t \subtxt{R})=g\Omega \subtxt{C}\exp\left(\mathrm{i}\bar{\omega} \subtxt{L}t \subtxt{R}+\frac{\xi}{2z \subtxt{g}}t \subtxt{R}^{2}\right)\times\\
\mathcal{F}_{z\rightarrow k_{z}=-\beta t \subtxt{R}}\left\{ \exp\left(\frac{\mathrm{i} zk_{\perp}^{2}}{2k \subtxt{0}}\right)\mathcal{\mathcal{F}}_{x,y\rightarrow k_{x},k_{y}}\left\{ \mathcal{S}(x,y,z)\right\} \right\} ,
\end{multline}
where we explicitly write the dominant part of the $\mathcal{O}(z,t)$
GEM phase correction which is a result of the slow decay of the magnetic
field gradient $\beta(t)=\beta \subtxt{0}-\xi t$ with $\xi t \subtxt{R}\ll\beta \subtxt{0}$.

Now the above relation is reversible. In the experiment, we use spatial heterodyne
detection to measure the amplitude and phase of the read-out light $\Omega \subtxt{s}$
in the far field for each value of $t \subtxt{R}$, i.e. exactly the left-hand side of the above equation. To recover the atomic part $\mathcal{S}(x,y,z)$ a series of multiplications and inverse Fourier transform is simply applied. Proper operation of the algorithm requires calibration of both the essential (such as the gradient) as well as nuisance (such as the rate of gradient decay) parameters.

\subsection*{Implementation}

The base of the experiment is a pencil-shaped ($10\times0.3\times0.3\ \mathrm{mm}^{3}$)
cloud of $^{\mathrm{87}}$Rb atoms formed in a magneto-optical trap (MOT) placed
in a constant magnetic field along $z$-axis: $\mathbf{B}=\hat{z}B \subtxt{0}$,
with \textbf{$B \subtxt{0}\approx\dimv{1}{G}$}. Atoms released from the
trap are optically pumped to $|\text{g}\rangle=5\text{S} \subtxt{1/2},\text{F}=2,\text{m} \subtxt{F}=2$ state, where $\text{F}$ denotes the total angular momentum and $\text{m} \subtxt{F}$ denotes a projection of the total angular momentum onto quantization ($z$) axis.

After pumping, a strong atomic coherence $\varrho \subtxt{hg}$ between $|\text{g}\rangle$
and $|\text{h}\rangle=5\text{S} \subtxt{1/2},\text{F}=1,\text{m} \subtxt{F}=0$ states is generated in a $\Lambda$
scheme with excited state $|\text{e}\rangle=5\text{P}\subtxt{1/2},\text{F}=2,\text{m} \subtxt{F}=1$, coupled
with mutually coherent input signal pulse (at $|\text{g}\rangle\rightarrow|\text{e}\rangle$
transition) and the control beam (at $|\text{e}\rangle\rightarrow|\text{h}\rangle$
transition), as shown in Fig. \ref{fig:Setup}\subfig{a}.

\begin{figure}
\includegraphics[width=1\columnwidth]{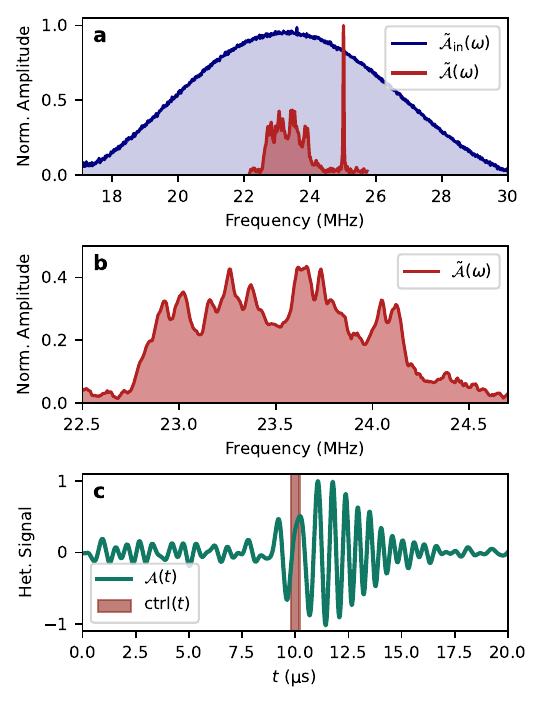}\caption{\textbf{Gradient Echo Memory (GEM) and input pulse spectra.} \subfig{A} Amplitude of \dimv{200}{ns} write-in pulse
(blue) and read-out signal (red) measured with heterodyne detection.
The spectrum of the write-in signal is much wider than the bandwidth
of the atomic cloud. The Zeeman splitting between the two-photon clock
transition ($\text{m} \subtxt{F}=1\to \text{m} \subtxt{F}=-1$, the narrow peak) and the memory transition ($\text{m} \subtxt{F}=2\to \text{m} \subtxt{F}=0$),
caused by the magnetic field bias, is also visible. \subfig{B} Zoomed-in spectrum of the read-out signal from panel \subfig{A}. \subfig{c} Temporal shape of the signal pulse registered by a point heterodyne (DPD). The red-shaded region represents the control pulse gating in the camera measurement.\label{fig:dpd}}
\end{figure}

To benchmark the tomography protocol we generate a flat atomic coherence
across the whole atomic ensemble by writing in a very short $200\ \mathrm{ns}$
input signal pulse accompanied by a control pulse of the same duration.
Such a short pulse populates spin waves along the entire length of
the cloud evenly, as the bandwidth of the input signal pulse is
much larger than the magnetically (inhomogeneously) broadened two-photon
absorption spectrum of the cloud. This was verified in measurement
shown in Figure \ref{fig:dpd}\subfig{A} where we compare the Fourier
magnitude of heterodyne detected input and output signals. Here,
we used a single point, differential photodiode (DPD) detector. The
measurement yields bandwidths of about $\dimv{8}{MHz}$ for the input
signal and about $\dimv{1.4}{MHz}$ for the registered output signal
that directly corresponds to the GEM bandwidth induced by the Zeeman
splitting gradient of around $2\uppi\times\dimv{1.4}{MHz\ cm^{-1}}$. The Fourier magnitude of the input signal is flat across the GEM spectrum
which guarantees the creation of a flat spin wave. The narrow peak visible around \dimv{25}{MHz} corresponds to the two-photon clock transition $\text{m} \subtxt{F}=1\to \text{m} \subtxt{F}=-1$ that due to the same Zeeman shift for the ground state sublevels is not affected by the magnetic field. Moreover, in the Figure
\ref{fig:dpd}\subfig{B} we show zoomed-in plot of the Fourier-magnitude
of the read-out signal that corresponds to the ($x$, $y$)-averaged
atomic density along the $z$-axis.

For the full 3D reconstruction of $\mathcal{S}(x,y,z)$ we replace
the DPD with an sCMOS camera located in the far field of the ensemble
with an effective focal length of about \dimv{250}{mm}. The two components of
the heterodyne optical signal are registered on two separate regions
of the camera that are then precisely aligned and subtracted to yield
differential images. The camera has a very limited temporal resolution
(exposure time of the order of 1 ms corresponds to 1 kHz bandwidth)
which would spoil the $z$-resolution. However, in GEM the temporal structure
of the read-out signal can be probed using a very short control pulse, as illustrated in Fig. \ref{fig:dpd}\subfig{c} where we show the temporal shape of the typical heterodyne trace registered by the DPD with the sampling period represented by ht red shaded region.
The full 3D spin wave tomography is then realized in a sequence
of measurements. In a single step, the control is turned on for only
$200\ \mathrm{ns}$, during which the read-out signal is generated
and registered by the camera. For each probing time $t \subtxt{R}$ we
collect 100 frames that are Fourier-filtered and coherently averaged
in real time. The filtering and averaging strongly decrease the
noise that is not correlated with the signal. The Fourier filtering
step in this coherent detection process corresponds to only taking
into account the signal which originated from the atoms and neglecting
the components that could not be created within the cloud transverse area. Then, to reconstruct the entire $\mathcal{S}(x,y,z)$, we register read-out
signal for 600 distinct control pulse delays at a step of \dimv{100}{ns}.
Finally, we get a three-dimensional array, which is a full Fourier transform
of the spin waves stored in the atomic ensemble. 

\subsection*{Calibration}
\begin{figure}
\includegraphics[width=1\columnwidth]{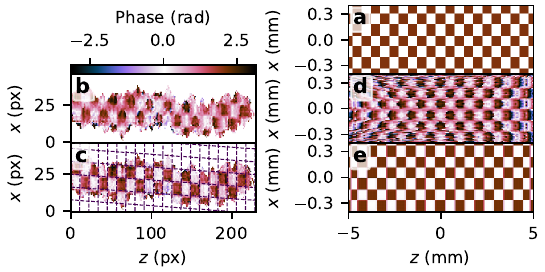}\caption{\textbf{Calibration and compensation.} \subfig{A} Checkerboard phase profile used for calibration imprinted onto atomic coherence before reading out the signal. \subfig{B}
The phase of raw three-dimensional Fourier transform of the read-out signal.
\subfig{C} The phase of the three-dimensional Fourier transform of the read-out
signal with compensation of both diffraction and temporal phase. The right column (\subfig{D, E}) corresponds to a numerical simulation of
the results from the left column (\subfig{B, C}). All panels share the same color scale.\label{fig:szachownica}}
\end{figure}

The correct 3D reconstruction of the spin wave phase and amplitude
requires calibration of the heterodyne camera setup. To benchmark and calibrate our device we use ac-Stark spatial spin wave
phase modulator (SSM). Using spatial light modulator
(SLM) and the camera, we are able to prepare a beam with arbitrarily chosen
intensity distribution, which illuminates atoms for $3\ \upmu\mathrm{s}$ after creating the coherence. Thanks to the ac-Stark effect, the atomic coherence
gets an additional phase which is proportional to the intensity of illuminating
light. A detailed description of this technique can be found in the works that utilize SSM for various protocols \cite{Lipka2019,Mazelanik2019,Parniak2019}. 

Figure \ref{fig:szachownica}\subfig{A}
presents a checkerboard intensity profile of the ac-Stark beam, which
we use for calibration. Figure \ref{fig:szachownica}\subfig{B} shows the phase cross-section
of the full Fourier transform of the signal registered by the camera
(without any compensation). In Fig. \ref{fig:szachownica}\subfig{d} we show a corresponding result of numerical simulation. In both cases, the checkerboard can be recognized, but
the retrieved image is blurred. In order to obtain a sharp pattern, two phenomena must be
taken into account. The first one is diffraction which manifests as
an additional quadratic phase in a Fourier domain that each slice attains
during propagation. The phase is $(z-z \subtxt{0})(k_{x}^{2}+k_{y}^{2})/(2k \subtxt{0})$,
where $k \subtxt{0}$ is the wave number of emitted signal and $k_{x,y}$
are transverse components of the wave vector. In real space, this results
in the blurred retrieved distribution in ($x$, $y$) subspace. The second
phenomenon is caused by a slow decrease in magnetic field gradient
strength during the read-out process. This effectively chirps the output
signal. In other words, the signal gets an additional quadratic phase
in the temporal domain, that results in a blur in $z$-direction. The
temporal phase to be compensated is $\zeta t^{2}$ with $\zeta=\dimv{-0.01}{rad\times \upmu s^{-2}}$.
Figure \ref{fig:szachownica}\subfig{C} (and numerical simulation in Fig. \ref{fig:szachownica}\subfig{e}) presents the retrieved phase checkerboard pattern after
both compensations. We finally see that the image is in focus and
sharp. The parameters ($z \subtxt{0}$ and $\zeta$) for optimal phase profiles
have been determined  to obtain the best sharpness of the resulting
images. Moreover, as the test pattern is prepared in real space
coordinates, the calibration procedure also yields the scaling factors
and rotation for $x,y,z$ axes. For this measurement, we extended the
measurement time window by collecting data for 1000 control pulse
delays spanning a range of $\dimv{100}{\upmu s}$. 
The absolute phase value is obtained by subtracting the reference (ref) image that was acquired
without the SSM modulation. The amplitude and the phase images are
masked to display only the points that correspond to an atomic coherence
magnitude much above the noise level $|\mathcal{S}^{\mathrm{ref}}(x,y,z)|>0.1$
and thus with a well-defined phase. 

\subsection*{Exemplary results}
\begin{figure}
\includegraphics[width=1\columnwidth]{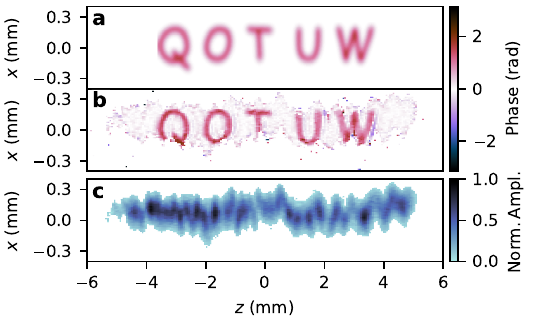}\caption{\textbf{Tomogram of atomic coherence.} \subfig{A} ``QOT UW'' inscription phase profile imprinted onto atomic coherence. \subfig{B} Phase of three-dimensional
Fourier transform of the measured signal with compensation of both
diffraction and temporal phase. \subfig{C} Amplitude of
fully compensated three-dimensional Fourier transform of the signal, which corresponds to atomic concentration.\label{fig:QOTUW}}
\end{figure}

With the calibration completed, we finally show some exemplary reconstructed phase and amplitude patterns of the atomic coherence. Figure \ref{fig:QOTUW} shows retrieved phase and amplitude profiles of a
flat spin wave phase modulated with an inscription ``QOT UW'' (abbreviation
of Quantum Optical Technologies, University of Warsaw). Figure \ref{fig:QOTUW}\subfig{A}
represents the displayed SSM pattern in phase (rad) units. In Fig.
\ref{fig:QOTUW}\subfig{B} we show the retrieved phase profile that
well matches the target profile. Additionally, in Fig. \ref{fig:QOTUW}\subfig{C}
we show retrieved spatial amplitude which in this case (flat spin
wave) corresponds to the atomic concentration. The full three-dimensional distribution of reconstructed phase and amplitude in the form of interactive visualization and a movie is available in Supplementary Data 1 and Supplementary Movie 1\cite{Mazelanik2023}.

\begin{figure}
\centering
\includegraphics{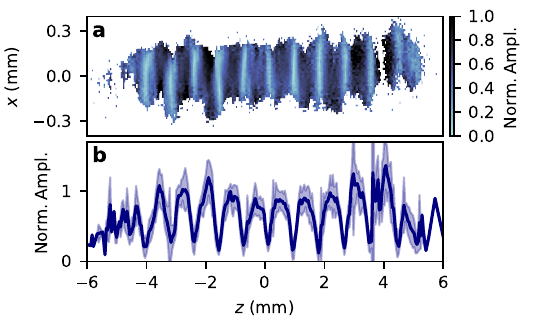}\caption{\textbf{Amplitude modulated coherence.} Slice of fully compensated three-dimensional Fourier transform of the signal corresponding to the coherence generated by two 200 ns input signal pulses separated by \dimv{8}{\upmu \mathrm{s}}. \subfig{A} Magnitude of two-dimensional $z$-$x$
slice of the normalized spin wave $\mathcal{S}(x,y,z)/\mathcal{S}^{\mathrm{ref}}(x,y,z)$.
\subfig{B} Average of the magnitude over $x$-axis, revealing the amplitude modulation. The shaded region represents the standard deviation of the sample's average. \label{fig:dudnienia}}
\end{figure}

To demonstrate the ability to reconstruct an amplitude pattern of
the spin wave we replace the flat spin wave with a modulated one.
This is accomplished by using two short ($200\ \mathrm{ns}$) input signal pulses ($200\ \mathrm{ns}$) separated by a delay of $\delta_{t}=8\ \mathrm{\upmu s}$.
The spectrum of two pulses of the same shape yet different amplitudes,
separated in time by $\delta_{t}$ is a phase and amplitude modulated
single pulse spectrum, given by the Fourier transform:
\begin{equation}
\mathcal{F}(\mathcal{A}(t)+\alpha\mathcal{A}(t+\delta_{t}))=\widetilde{\mathcal{A}}(\omega)(1+\alpha \mathrm{e}^{\mathrm{i}\omega\delta_{t}}).
\end{equation}
Due to the spectrum to position mapping feature of GEM \cite{Mazelanik2020,Mazelanik2022}
combined with the large input signal bandwidth (see Fig. \ref{fig:dpd}) this yields uniformly modulated coherence $\varrho \subtxt{hg}\propto(1+\alpha \mathrm{e}^{2 \uppi \mathrm{i} \delta_{t}\beta z})$.
Figure \ref{fig:dudnienia}\subfig{A} presents a two-dimensional magnitude
slice $|\varrho \subtxt{hg}(x,z)|$ of the retrieved and fully compensated
spin wave pattern normalized to a flat spin wave:
$\varrho \subtxt{hg}(x,y,z)=\mathcal{S}(x,y,z)/\mathcal{S}^{\mathrm{ref}}(x,y,z)$.
The magnitude, as expected, resembles a modulus of cosine function $|\varrho \subtxt{hg}(x,z)|\propto|\cos(2 \uppi \delta t\beta z)|$
as the relative amplitude ratio is close to unity $\alpha\approx1$.
Figure \ref{fig:dudnienia}\subfig{B} represents the normalized absolute value of the coherence slice
$|\varrho \subtxt{hg}(x,z)|$ averaged over the $x$-axis.

Let us now shortly discuss the resolution limitation of the demonstrated 3D imaging method.
The resolution in the perpendicular coordinates $(x,y)$ is limited
by the optical imaging system, namely its numerical aperture. From
the calibration image in Fig. \ref{fig:szachownica}\subfig{c} we can estimate
the resolution as the half-width of the phase slope, which amounts
to about \dimv{1}{px}. That equals to $\delta x,\delta y\approx\dimv{14}{\upmu m}$.
The resolution along the propagation axis is however limited by the
duration of the measurement window $T_{\mathrm{meas}}=\dimv{100}{\upmu s}$ that yields the minimal width of a feature in the spectral domain $\delta\omega\simeq2\uppi\times1/T$, 
combined with the magnitude of the magnetic field gradient $\mu \subtxt{B}\beta/\hbar=2 \uppi \times\dimv{1.4}{MHz \ cm^{-1}}$ that facilitates the spectrum to position mapping $\mu \subtxt{B} \beta z/\hbar \leftrightarrow  \omega$.
The longitudinal resolution could be thus estimated by taking half
of the inverse of the product of the measurement window and Zeeman splitting
gradient: $\delta z\simeq0.5/(T_{\mathrm{meas}}\beta\mu \subtxt{B}/\hbar)\approx\dimv{36}{\upmu m}$.
In the calibration images (Fig. \ref{fig:szachownica}\subfig{C})
we see that the width of the phase slope in the $z$-direction spans
approximately \dimv{2}{px}. This yields $\dimv{44}{\upmu m}$ of resolution.
The maximal duration of the measurement window is in fact limited by the
thermal motion of the atoms inside the ensemble \cite{Parniak2017, Lipka2021}. By measuring the output signal intensity for a range of storage times (time between creation and retrieval of the coherence) we estimate that after $\dimv{120}{\upmu s}$ the read-out efficiency drops by $50\%$, which corresponds to an ensemble with a temperature of $T=\dimv{265\pm15}{\upmu K}$ (see Methods). This is significantly larger than a typical ultracold system such as a Bose-Einstein condensate (with Rb atoms), which demonstrates that our method does not require extreme levels of optical cooling, and performs excellently with only roughly Doppler-limited cooling. 

\begin{figure}
\centering
\includegraphics[width=1\columnwidth]{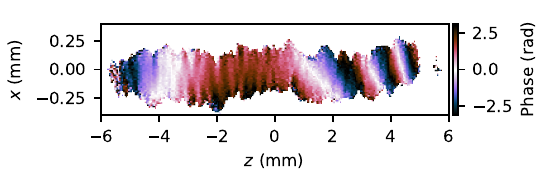}

\caption{\textbf{3D magnetometry.} Retrieved phase (see colorbar) of the coherence $\varrho \subtxt{hg}(x,z)$ affected by
an inhomogeneous magnetic field generated by a small coil placed near
the ensemble. \label{fig:magneto} }
\end{figure}

Finally, we demonstrate the potential for application in 3D magnetometry
by detecting a spin wave phase modulation caused by the magnetic field
generated by a small coil placed above the atomic ensemble (outside the vacuum chamber). The coil is turned on for a short period in the sequence after the creation of the coherence. The inhomogeneous magnetic field generated by
the coil imprints a phase that is proportional to the total magnitude
of the magnetic field $|\mathbf{B}|$ and the interaction time $t \subtxt{c}$.
The accumulated phase is $\varphi=\mu \subtxt{B}|\vect{B}+\hat{z}B \subtxt{0}|t \subtxt{c}/\hbar$.
The presence of the constant magnetic field along the $z$-axis makes
the phase sensitive to mostly the changes of the field along the $z$-axis.
However, one could easily imagine a more advanced protocol in which
for the time of the measurement the constant field component is switched
off. In such a scenario additionally to the phase modulation, one should
observe amplitude modulation caused by the atomic spin rotation that
yields different projections onto the $z$-axis and modifies the read-out process
efficiency (the coherence is partially transferred to a different
magnetic sublevel, that is not coupled by the control laser). Figure
\ref{fig:magneto} shows the retrieved phase profile of the magnetically
modulated coherence. The constant phase value lines correspond to
the constant values of the local magnetic field. 

\section*{Conclusions}

We have demonstrated how to reconstruct a 3D complex spatial distribution
of atomic coherence stored in the atomic ensemble. Our method employs a magnetic field gradient to map the longitudinal components of the coherence
onto the optical signal frequency. A single spatiotemporally resolved
heterodyne measurement of the signal light allows reconstruction
of the full complex distribution of the atomic coherence $S(x,y,z)=n(x,y,z)\rohg(x,y,z)$. The unwinding of phase accumulated due to diffraction and compensating for distortions enables the faithful reconstruction of the complex atomic coherence.

Here, for the detection, we use a time-gated camera that requires many
(short) measurements corresponding to different components in the
reciprocal space of the propagation axis. However, a 2D array of
photodiodes coupled with fast analog to digital converters would allow
a single-shot measurement of the full 3D distribution. Moreover, a
much simpler 1D array could be used in a hybrid measurement scheme
with one of the axes resolved using a rainbow heterodyning technique \cite{Strauss1994}
enabled by a multi-frequency LO with frequency gradient along the
given (orthogonal to the array) axis. Finally, the demonstrated 3D
magnetometry protocol could be extended to enable a 3D tomography of electric
and electromagnetic fields in the microwave regime \cite{Sedlacek2012}, opening a new
kind of atom-based metrology. Although our demonstration covers a millimeter-scale cold ensemble the protocol could be implemented in larger systems such as glass cells containing warm atoms, enabling centimeter-scale tomography. Beyond measuring external influences, the three-dimensional structure of Rydberg-atom interactions and propagation of Rydberg polaritons could be interrogated with the presented method, in order to generate exotic states of matter such as Efimov states \cite{Gullans2017} efficiently.

\section*{Methods}
\subsection*{Quantum Memory setup}
The GEM is based on a cold rubidium-87 ensemble prepared in a magnetooptical trap (MOT). The MOT utilizes quadruple coils and 3 pairs of counter-propagating trapping and cooling beams, which generate a strong symmetric trapping potential in $x,y$ axes. The trapping potential along the propagation ($z$) axis is much weaker. The resulting ensemble is thus elongated in the $z$-direction with approximate dimensions of $10\times0.3\times0.3\ \mathrm{mm}^{3}$. The trap is held in a bias magnetic field along $z$-axis of $\mathbf{B}=\hat{z}B \subtxt{0}$,
with \textbf{$B \subtxt{0}\approx\dimv{1}{G}$}, that cancels the magnetic sublevels degeneration. After the trapping period (ca. $\dimv{20}{ms}$) we optically pump the atoms to to $|\text{g}\rangle=5\text{S} \subtxt{1/2},\text{F}=2,\text{m} \subtxt{F}=2$ state. This is achieved by illuminating the cloud with two laser beams 
for $\dimv{15}{\upmu s}$. The first beam at
\dimv{795}{nm} is tuned to $5\text{S} \subtxt{1/2},\text{F}=1\rightarrow5\text{P}\subtxt{1/2},\text{F}=2$
transition and illuminates the cloud from 4 sides. The beam provides hyperfine pumping by emptying the $5\text{S} \subtxt{1/2},\text{F}=1$ ground level. The second $\sigma_{+}$ circularly polarized beam at \dimv{780}{nm} is tuned to $5\text{S} \subtxt{1/2},\text{F}=2\rightarrow5\text{P} \subtxt{3/2},\text{F}=2$
transition and illuminates the atoms along the propagation axis, providing magnetic sublevel pumping within the $5\text{S}\subtxt{1/2},\text{F}=2$
manifold, ideally populating only the $|\text{g}\rangle$ state.
The GEM coils are square-shaped with \dimv{10}{cm} side length, and are separated by \dimv{18}{cm}. They produce a linear magnetic field gradient along the cloud of approximately $\dimv{1}{G \ cm^{-1}}$, which can be rapidly ($\dimv{0.35}{G\ cm^{-1} \ \upmu s^{-1}}$) switched to the opposite.

\subsection*{Heterodyne camera}
The heterodyne camera setup incorporates a single Scientific-CMOS (sCMOS) camera (Andor Zyla) located in the far field of the atomic ensemble. 
The camera pane is divided into two regions (two ports of the heterodyne) that are precisely aligned and subtracted during measurements. The local oscillator (LO) is derived from the same laser as the control and signal fields, and is shifted in frequency by around $\dimv{1}{MHz}$. Moreover the LO is slightly angled with respect to the signal field, which produces spatial interference fringes in the measurement (spatial heterodyne).  
The spatial and temporal frequency shifts allow us to separate the genuine signal from background noises. The differential heterodyne frames are coherently averaged for each measurement point. The averaging is possible thanks to additional reference frame preceding each signal frame. For this we set the camera in the Particle Imaging Velocimetry (PIV) mode which enables collection of two frames separated by very short delay ($\dimv{2}{\upmu s}$). The first frame collects light from the incoming signal pulse that is used to create atomic coherence, and provides a phase reference for the second frame containing the read-out signal. For the coherent averaging we first unwind the phase of each signal frame according to the reference frame and then simply sum all the signal frames for a given measurement time $t \subtxt{R}$. It is worth noting here that the phase fluctuations that we are compensating this way are solely caused by the signal-control-LO optical path length difference fluctuations, and not the laser itself (as all the beams are coming from the same laser, and the global phase fluctuations cancel out in the heterodyne measurement).

\subsection{Thermal decoherence}

\begin{figure}[t]
\centering
\includegraphics[width=0.9\columnwidth]{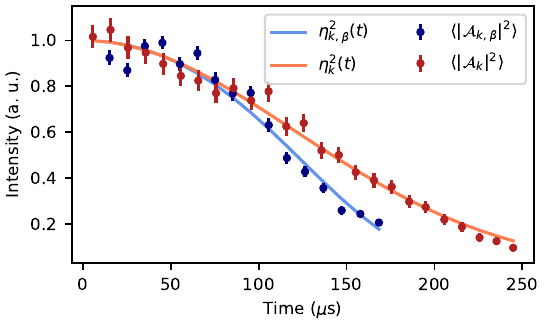}
\caption{\textbf{Thermal decoherence of the spin waves in Gradient Echo Memory (GEM).} The data (points) represents the decay of the retrieved signal field without ($\langle|\mathcal{A}_{\mathrm{k}}|^2\rangle$, red) and with GEM protocol ($\langle|\mathcal{A}_{\mathrm{k}, \upbeta}|^2\rangle$, blue). The solid lines correspond to fitted theoretical curves $\eta_{\mathrm{k}}^2$ (orange) and $\eta_{\mathrm{k}, \upbeta}^2$ (blue). Measured lifetimes are
equal $\tau_{\mathrm{k}}\approx\dimv{173}{\upmu s}$ and $\tau_{\upbeta}\approx\dimv{175}{\upmu s}$,
which corresponds to $\dimv{265\pm15}{\upmu K}$. The error bars represent the standard deviation of the data. \label{fig:temperatura}}
\end{figure}

To estimate the rate of thermal decoherence caused by the atomic motion, we measured the output
signal amplitude for a range of read-out times $t\subtxt{R}$. The
results are presented in Figure \ref{fig:temperatura}. The measurements
were performed with and without magnetic field gradient. The thermal
decoherence of spin waves in an atomic ensemble without the magnetic
field gradient is well known and yields a Gaussian decay \cite{Parniak2017}.
In the GEM, due to the magnetic field gradient, atoms traveling along the ensemble
enter regions with different values of the magnetic field and attain
additional phase. Namely, each group of atoms with velocity $v_{z}$
gains an additional phase factor of $\exp(\mathrm{i}\beta v_{z}t\subtxt{R}^{2}/4)$.
That, after averaging with Maxwell velocity distribution, gives additional
exponential decoherence term with time in the fourth power. The full
expression for the read-out decay reads:
\begin{equation}
\begin{gathered}\mathcal{\eta}_{\mathrm{k},\upbeta}(t)=\mathrm{e}^{-\frac{t^{2}}{2\tau_{\mathrm{k}}^{2}}-\frac{t^{4}}{2\tau_{\upbeta}^{4}}},\\
\tau_{\mathrm{k}}=\frac{1}{k_{\mathcal{S}}}\sqrt{\frac{m}{k_{\mathrm{b}}T}}\text{, \ensuremath{\tau_{\upbeta}}=\ensuremath{\sqrt{\frac{2}{\uppi\beta}\sqrt{\frac{m}{k_{\mathrm{b}}T}}}}},\\
\\
\end{gathered}
\end{equation}
where $k_{\mathcal{S}}$ is the spin wave wavevector, $k_{\mathrm{b}}$ is Boltzmann
constant, $m$ is $^{87}$Rb mass and $T$ is temperature of atomic
cloud. The solid curves in Fig. \ref{fig:temperatura} correspond
to the above model fitted to the experimental data. From the measurement
without magnetic field gradient, we recover the characteristic time
$\ensuremath{\tau_{\mathrm{k}}=\dimv{173\pm5}{\upmu s}}$. This for the angle between the control and
the write-in signal beam of $4.6\ \mathrm{mrad}$ corresponds to a temperature of $T=\dimv{265\pm15}{\upmu K}$.
The second measurement with the gradient turned on yields $\ensuremath{\tau_{\upbeta}=\dimv{175.4\pm2.5}{\upmu s}}$.
To compensate for this decay in the reconstruction procedure we divide
the measured signal $\mathcal{A}(t \subtxt{R})$ by the factor $\eta_{\mathrm{k},\upbeta}(t\subtxt{R})$.

\section*{Data Availability}
The data presented in the figures of this manuscript has been deposited at \cite{dane}.

\section*{Acknowledgments}
We thank K. Banaszek, W. Wasilewski for their support and K. Jachymski and S. Borówka for insightful discussions.
The “Quantum Optical Technologies” (MAB/2018/4) project is carried
out within the International Research Agendas programme of the Foundation
for Polish Science co-financed by the European Union under the European
Regional Development Fund. MM was also supported by the Foundation
for Polish Science via the START scholarship.  This research was funded
in whole or in part by National Science Centre, Poland grant no. 2021/43/D/ST2/03114
and by the Office of Naval Research Global grant no. N62909-19-1-2127.

\section*{Author contributions}
MM and MP conceived the protocol and experimental implementation. MM, AL, and MP contributed to the experimental setup, and data collection. MM analyzed the data and prepared figures with help from AL. MM, AL, and MP developed the theory for which TS and AL performed the numerical simulations. MM, AL and MP wrote the manuscript with contributions from TS. MM and MP managed the project.
\section*{Competing Interests}
The authors declare no competing interests.

\bibliography{bib}

\begin{thebibliography}{47}%
\makeatletter
\providecommand \@ifxundefined [1]{%
 \@ifx{#1\undefined}
}%
\providecommand \@ifnum [1]{%
 \ifnum #1\expandafter \@firstoftwo
 \else \expandafter \@secondoftwo
 \fi
}%
\providecommand \@ifx [1]{%
 \ifx #1\expandafter \@firstoftwo
 \else \expandafter \@secondoftwo
 \fi
}%
\providecommand \natexlab [1]{#1}%
\providecommand \enquote  [1]{``#1''}%
\providecommand \bibnamefont  [1]{#1}%
\providecommand \bibfnamefont [1]{#1}%
\providecommand \citenamefont [1]{#1}%
\providecommand \href@noop [0]{\@secondoftwo}%
\providecommand \href [0]{\begingroup \@sanitize@url \@href}%
\providecommand \@href[1]{\@@startlink{#1}\@@href}%
\providecommand \@@href[1]{\endgroup#1\@@endlink}%
\providecommand \@sanitize@url [0]{\catcode `\\12\catcode `\$12\catcode
  `\&12\catcode `\#12\catcode `\^12\catcode `\_12\catcode `\%12\relax}%
\providecommand \@@startlink[1]{}%
\providecommand \@@endlink[0]{}%
\providecommand \url  [0]{\begingroup\@sanitize@url \@url }%
\providecommand \@url [1]{\endgroup\@href {#1}{\urlprefix }}%
\providecommand \urlprefix  [0]{URL }%
\providecommand \Eprint [0]{\href }%
\providecommand \doibase [0]{https://doi.org/}%
\providecommand \selectlanguage [0]{\@gobble}%
\providecommand \bibinfo  [0]{\@secondoftwo}%
\providecommand \bibfield  [0]{\@secondoftwo}%
\providecommand \translation [1]{[#1]}%
\providecommand \BibitemOpen [0]{}%
\providecommand \bibitemStop [0]{}%
\providecommand \bibitemNoStop [0]{.\EOS\space}%
\providecommand \EOS [0]{\spacefactor3000\relax}%
\providecommand \BibitemShut  [1]{\csname bibitem#1\endcsname}%
\let\auto@bib@innerbib\@empty
\bibitem [{\citenamefont {Shin}\ \emph {et~al.}(2022)\citenamefont {Shin},
  \citenamefont {Eun}, \citenamefont {Lee}, \citenamefont {Lee}, \citenamefont
  {Hugonnet}, \citenamefont {Yoon}, \citenamefont {Kim}, \citenamefont
  {Jeong},\ and\ \citenamefont {Park}}]{Shin2022}%
  \BibitemOpen
  \bibfield  {author} {\bibinfo {author} {\bibfnamefont {S.}~\bibnamefont
  {Shin}}, \bibinfo {author} {\bibfnamefont {J.}~\bibnamefont {Eun}}, \bibinfo
  {author} {\bibfnamefont {S.~S.}\ \bibnamefont {Lee}}, \bibinfo {author}
  {\bibfnamefont {C.}~\bibnamefont {Lee}}, \bibinfo {author} {\bibfnamefont
  {H.}~\bibnamefont {Hugonnet}}, \bibinfo {author} {\bibfnamefont {D.~K.}\
  \bibnamefont {Yoon}}, \bibinfo {author} {\bibfnamefont {S.-H.}\ \bibnamefont
  {Kim}}, \bibinfo {author} {\bibfnamefont {J.}~\bibnamefont {Jeong}},\ and\
  \bibinfo {author} {\bibfnamefont {Y.}~\bibnamefont {Park}},\ }\bibfield
  {title} {{\selectlanguage {en}\bibinfo {title} {Tomographic measurement of
  dielectric tensors at optical frequency}},\ }\href
  {https://doi.org/10.1038/s41563-022-01202-8} {\bibfield  {journal} {\bibinfo
  {journal} {Nature Materials}\ }\textbf {\bibinfo {volume} {21}},\ \bibinfo
  {pages} {317} (\bibinfo {year} {2022})}\BibitemShut {NoStop}%
\bibitem [{\citenamefont {Engström}\ \emph {et~al.}(2011)\citenamefont
  {Engström}, \citenamefont {Trivedi}, \citenamefont {Persson}, \citenamefont
  {Goksör}, \citenamefont {Bertness},\ and\ \citenamefont
  {Smalyukh}}]{Engstroem2011}%
  \BibitemOpen
  \bibfield  {author} {\bibinfo {author} {\bibfnamefont {D.}~\bibnamefont
  {Engström}}, \bibinfo {author} {\bibfnamefont {R.~P.}\ \bibnamefont
  {Trivedi}}, \bibinfo {author} {\bibfnamefont {M.}~\bibnamefont {Persson}},
  \bibinfo {author} {\bibfnamefont {M.}~\bibnamefont {Goksör}}, \bibinfo
  {author} {\bibfnamefont {K.~A.}\ \bibnamefont {Bertness}},\ and\ \bibinfo
  {author} {\bibfnamefont {I.~I.}\ \bibnamefont {Smalyukh}},\ }\bibfield
  {title} {{\selectlanguage {en}\bibinfo {title} {Three-dimensional imaging of
  liquid crystal structures and defects by means of holographic manipulation of
  colloidal nanowires with faceted sidewalls}},\ }\href
  {https://doi.org/10.1039/C1SM05170A} {\bibfield  {journal} {\bibinfo
  {journal} {Soft Matter}\ }\textbf {\bibinfo {volume} {7}},\ \bibinfo {pages}
  {6304} (\bibinfo {year} {2011})}\BibitemShut {NoStop}%
\bibitem [{\citenamefont {Sung}(2021)}]{Sung2021}%
  \BibitemOpen
  \bibfield  {author} {\bibinfo {author} {\bibfnamefont {Y.}~\bibnamefont
  {Sung}},\ }\bibfield  {title} {\bibinfo {title} {Snapshot
  {Three}-{Dimensional} {Absorption} {Imaging} of {Microscopic} {Specimens}},\
  }\href {https://doi.org/10.1103/PhysRevApplied.15.064065} {\bibfield
  {journal} {\bibinfo  {journal} {Physical Review Applied}\ }\textbf {\bibinfo
  {volume} {15}},\ \bibinfo {pages} {064065} (\bibinfo {year}
  {2021})}\BibitemShut {NoStop}%
\bibitem [{\citenamefont {Xiao}\ \emph {et~al.}(2013)\citenamefont {Xiao},
  \citenamefont {Javidi}, \citenamefont {Martinez-Corral},\ and\ \citenamefont
  {Stern}}]{Xiao2013}%
  \BibitemOpen
  \bibfield  {author} {\bibinfo {author} {\bibfnamefont {X.}~\bibnamefont
  {Xiao}}, \bibinfo {author} {\bibfnamefont {B.}~\bibnamefont {Javidi}},
  \bibinfo {author} {\bibfnamefont {M.}~\bibnamefont {Martinez-Corral}},\ and\
  \bibinfo {author} {\bibfnamefont {A.}~\bibnamefont {Stern}},\ }\bibfield
  {title} {{\selectlanguage {EN}\bibinfo {title} {Advances in three-dimensional
  integral imaging: sensing, display, and applications [{Invited}]}},\ }\href
  {https://doi.org/10.1364/AO.52.000546} {\bibfield  {journal} {\bibinfo
  {journal} {Applied Optics}\ }\textbf {\bibinfo {volume} {52}},\ \bibinfo
  {pages} {546} (\bibinfo {year} {2013})}\BibitemShut {NoStop}%
\bibitem [{\citenamefont {Lauterbur}(1973)}]{Lauterbur1973}%
  \BibitemOpen
  \bibfield  {author} {\bibinfo {author} {\bibfnamefont {P.~C.}\ \bibnamefont
  {Lauterbur}},\ }\bibfield  {title} {\bibinfo {title} {Image formation by
  induced local interactions: Examples employing nuclear magnetic resonance},\
  }\href {https://doi.org/10.1038/242190a0} {\bibfield  {journal} {\bibinfo
  {journal} {Nature}\ }\textbf {\bibinfo {volume} {242}},\ \bibinfo {pages}
  {190} (\bibinfo {year} {1973})}\BibitemShut {NoStop}%
\bibitem [{\citenamefont {Feinberg}\ \emph {et~al.}(2010)\citenamefont
  {Feinberg}, \citenamefont {Moeller}, \citenamefont {Smith}, \citenamefont
  {Auerbach}, \citenamefont {Ramanna}, \citenamefont {Glasser}, \citenamefont
  {Miller}, \citenamefont {Ugurbil},\ and\ \citenamefont
  {Yacoub}}]{Feinberg2010}%
  \BibitemOpen
  \bibfield  {author} {\bibinfo {author} {\bibfnamefont {D.~A.}\ \bibnamefont
  {Feinberg}}, \bibinfo {author} {\bibfnamefont {S.}~\bibnamefont {Moeller}},
  \bibinfo {author} {\bibfnamefont {S.~M.}\ \bibnamefont {Smith}}, \bibinfo
  {author} {\bibfnamefont {E.}~\bibnamefont {Auerbach}}, \bibinfo {author}
  {\bibfnamefont {S.}~\bibnamefont {Ramanna}}, \bibinfo {author} {\bibfnamefont
  {M.~F.}\ \bibnamefont {Glasser}}, \bibinfo {author} {\bibfnamefont {K.~L.}\
  \bibnamefont {Miller}}, \bibinfo {author} {\bibfnamefont {K.}~\bibnamefont
  {Ugurbil}},\ and\ \bibinfo {author} {\bibfnamefont {E.}~\bibnamefont
  {Yacoub}},\ }\bibfield  {title} {\bibinfo {title} {Multiplexed echo planar
  imaging for sub-second whole brain fmri and fast diffusion imaging},\ }\href
  {https://doi.org/10.1371/journal.pone.0015710} {\bibfield  {journal}
  {\bibinfo  {journal} {PLOS ONE}\ }\textbf {\bibinfo {volume} {5}},\ \bibinfo
  {pages} {e15710} (\bibinfo {year} {2010})}\BibitemShut {NoStop}%
\bibitem [{\citenamefont {Allodi}\ \emph {et~al.}(2016)\citenamefont {Allodi},
  \citenamefont {Dahlberg}, \citenamefont {Mazuski}, \citenamefont {Davis},
  \citenamefont {Otto},\ and\ \citenamefont {Engel}}]{Allodi2016}%
  \BibitemOpen
  \bibfield  {author} {\bibinfo {author} {\bibfnamefont {M.~A.}\ \bibnamefont
  {Allodi}}, \bibinfo {author} {\bibfnamefont {P.~D.}\ \bibnamefont
  {Dahlberg}}, \bibinfo {author} {\bibfnamefont {R.~J.}\ \bibnamefont
  {Mazuski}}, \bibinfo {author} {\bibfnamefont {H.~C.}\ \bibnamefont {Davis}},
  \bibinfo {author} {\bibfnamefont {J.~P.}\ \bibnamefont {Otto}},\ and\
  \bibinfo {author} {\bibfnamefont {G.~S.}\ \bibnamefont {Engel}},\ }\bibfield
  {title} {\bibinfo {title} {Optical resonance imaging: An optical analog to
  mri with subdiffraction-limited capabilities},\ }\href
  {https://doi.org/10.1021/acsphotonics.6b00694} {\bibfield  {journal}
  {\bibinfo  {journal} {ACS Photonics}\ }\textbf {\bibinfo {volume} {3}},\
  \bibinfo {pages} {2445} (\bibinfo {year} {2016})}\BibitemShut {NoStop}%
\bibitem [{\citenamefont {Kasai}\ \emph {et~al.}(2018)\citenamefont {Kasai},
  \citenamefont {Okamoto}, \citenamefont {Nishioka}, \citenamefont {Takagi},\
  and\ \citenamefont {Sasaki}}]{Kasai2018}%
  \BibitemOpen
  \bibfield  {author} {\bibinfo {author} {\bibfnamefont {J.}~\bibnamefont
  {Kasai}}, \bibinfo {author} {\bibfnamefont {Y.}~\bibnamefont {Okamoto}},
  \bibinfo {author} {\bibfnamefont {K.}~\bibnamefont {Nishioka}}, \bibinfo
  {author} {\bibfnamefont {T.}~\bibnamefont {Takagi}},\ and\ \bibinfo {author}
  {\bibfnamefont {Y.}~\bibnamefont {Sasaki}},\ }\bibfield  {title} {\bibinfo
  {title} {Chiral {Domain} {Structure} in {Superfluid} $^3\mathrm{He}$-${A}$
  {Studied} by {Magnetic} {Resonance} {Imaging}},\ }\href
  {https://doi.org/10.1103/PhysRevLett.120.205301} {\bibfield  {journal}
  {\bibinfo  {journal} {Physical Review Letters}\ }\textbf {\bibinfo {volume}
  {120}},\ \bibinfo {pages} {205301} (\bibinfo {year} {2018})}\BibitemShut
  {NoStop}%
\bibitem [{\citenamefont {Karpov}\ \emph {et~al.}(2017)\citenamefont {Karpov},
  \citenamefont {Liu}, \citenamefont {Rolo}, \citenamefont {Harder},
  \citenamefont {Balachandran}, \citenamefont {Xue}, \citenamefont {Lookman},\
  and\ \citenamefont {Fohtung}}]{Karpov2017}%
  \BibitemOpen
  \bibfield  {author} {\bibinfo {author} {\bibfnamefont {D.}~\bibnamefont
  {Karpov}}, \bibinfo {author} {\bibfnamefont {Z.}~\bibnamefont {Liu}},
  \bibinfo {author} {\bibfnamefont {T.~d.~S.}\ \bibnamefont {Rolo}}, \bibinfo
  {author} {\bibfnamefont {R.}~\bibnamefont {Harder}}, \bibinfo {author}
  {\bibfnamefont {P.~V.}\ \bibnamefont {Balachandran}}, \bibinfo {author}
  {\bibfnamefont {D.}~\bibnamefont {Xue}}, \bibinfo {author} {\bibfnamefont
  {T.}~\bibnamefont {Lookman}},\ and\ \bibinfo {author} {\bibfnamefont
  {E.}~\bibnamefont {Fohtung}},\ }\bibfield  {title} {{\selectlanguage
  {en}\bibinfo {title} {Three-dimensional imaging of vortex structure in a
  ferroelectric nanoparticle driven by an electric field}},\ }\href
  {https://doi.org/10.1038/s41467-017-00318-9} {\bibfield  {journal} {\bibinfo
  {journal} {Nature Communications}\ }\textbf {\bibinfo {volume} {8}},\
  \bibinfo {pages} {280} (\bibinfo {year} {2017})}\BibitemShut {NoStop}%
\bibitem [{\citenamefont {Kardjilov}\ \emph {et~al.}(2008)\citenamefont
  {Kardjilov}, \citenamefont {Manke}, \citenamefont {Strobl}, \citenamefont
  {Hilger}, \citenamefont {Treimer}, \citenamefont {Meissner}, \citenamefont
  {Krist},\ and\ \citenamefont {Banhart}}]{Kardjilov2008}%
  \BibitemOpen
  \bibfield  {author} {\bibinfo {author} {\bibfnamefont {N.}~\bibnamefont
  {Kardjilov}}, \bibinfo {author} {\bibfnamefont {I.}~\bibnamefont {Manke}},
  \bibinfo {author} {\bibfnamefont {M.}~\bibnamefont {Strobl}}, \bibinfo
  {author} {\bibfnamefont {A.}~\bibnamefont {Hilger}}, \bibinfo {author}
  {\bibfnamefont {W.}~\bibnamefont {Treimer}}, \bibinfo {author} {\bibfnamefont
  {M.}~\bibnamefont {Meissner}}, \bibinfo {author} {\bibfnamefont
  {T.}~\bibnamefont {Krist}},\ and\ \bibinfo {author} {\bibfnamefont
  {J.}~\bibnamefont {Banhart}},\ }\bibfield  {title} {{\selectlanguage
  {en}\bibinfo {title} {Three-dimensional imaging of magnetic fields with
  polarized neutrons}},\ }\href {https://doi.org/10.1038/nphys912} {\bibfield
  {journal} {\bibinfo  {journal} {Nature Physics}\ }\textbf {\bibinfo {volume}
  {4}},\ \bibinfo {pages} {399} (\bibinfo {year} {2008})}\BibitemShut {NoStop}%
\bibitem [{\citenamefont {Kim}\ \emph {et~al.}(2018)\citenamefont {Kim},
  \citenamefont {Chamard}, \citenamefont {Hallmann}, \citenamefont {Roth},
  \citenamefont {Lu}, \citenamefont {Boesenberg}, \citenamefont {Zozulya},
  \citenamefont {Leake},\ and\ \citenamefont {Madsen}}]{Kim2018}%
  \BibitemOpen
  \bibfield  {author} {\bibinfo {author} {\bibfnamefont {C.}~\bibnamefont
  {Kim}}, \bibinfo {author} {\bibfnamefont {V.}~\bibnamefont {Chamard}},
  \bibinfo {author} {\bibfnamefont {J.}~\bibnamefont {Hallmann}}, \bibinfo
  {author} {\bibfnamefont {T.}~\bibnamefont {Roth}}, \bibinfo {author}
  {\bibfnamefont {W.}~\bibnamefont {Lu}}, \bibinfo {author} {\bibfnamefont
  {U.}~\bibnamefont {Boesenberg}}, \bibinfo {author} {\bibfnamefont
  {A.}~\bibnamefont {Zozulya}}, \bibinfo {author} {\bibfnamefont
  {S.}~\bibnamefont {Leake}},\ and\ \bibinfo {author} {\bibfnamefont
  {A.}~\bibnamefont {Madsen}},\ }\bibfield  {title} {\bibinfo {title}
  {Three-{Dimensional} {Imaging} of {Phase} {Ordering} in an {Fe}-{Al} {Alloy}
  by {Bragg} {Ptychography}},\ }\href
  {https://doi.org/10.1103/PhysRevLett.121.256101} {\bibfield  {journal}
  {\bibinfo  {journal} {Physical Review Letters}\ }\textbf {\bibinfo {volume}
  {121}},\ \bibinfo {pages} {256101} (\bibinfo {year} {2018})}\BibitemShut
  {NoStop}%
\bibitem [{\citenamefont {Rugar}\ \emph {et~al.}(2004)\citenamefont {Rugar},
  \citenamefont {Budakian}, \citenamefont {Mamin},\ and\ \citenamefont
  {Chui}}]{Rugar2004}%
  \BibitemOpen
  \bibfield  {author} {\bibinfo {author} {\bibfnamefont {D.}~\bibnamefont
  {Rugar}}, \bibinfo {author} {\bibfnamefont {R.}~\bibnamefont {Budakian}},
  \bibinfo {author} {\bibfnamefont {H.~J.}\ \bibnamefont {Mamin}},\ and\
  \bibinfo {author} {\bibfnamefont {B.~W.}\ \bibnamefont {Chui}},\ }\bibfield
  {title} {{\selectlanguage {en}\bibinfo {title} {Single spin detection by
  magnetic resonance force microscopy}},\ }\href
  {https://doi.org/10.1038/nature02658} {\bibfield  {journal} {\bibinfo
  {journal} {Nature}\ }\textbf {\bibinfo {volume} {430}},\ \bibinfo {pages}
  {329} (\bibinfo {year} {2004})}\BibitemShut {NoStop}%
\bibitem [{\citenamefont {Streed}\ \emph {et~al.}(2012)\citenamefont {Streed},
  \citenamefont {Jechow}, \citenamefont {Norton},\ and\ \citenamefont
  {Kielpinski}}]{Streed2012}%
  \BibitemOpen
  \bibfield  {author} {\bibinfo {author} {\bibfnamefont {E.~W.}\ \bibnamefont
  {Streed}}, \bibinfo {author} {\bibfnamefont {A.}~\bibnamefont {Jechow}},
  \bibinfo {author} {\bibfnamefont {B.~G.}\ \bibnamefont {Norton}},\ and\
  \bibinfo {author} {\bibfnamefont {D.}~\bibnamefont {Kielpinski}},\ }\bibfield
   {title} {{\selectlanguage {en}\bibinfo {title} {Absorption imaging of a
  single atom}},\ }\href {https://doi.org/10.1038/ncomms1944} {\bibfield
  {journal} {\bibinfo  {journal} {Nature Communications}\ }\textbf {\bibinfo
  {volume} {3}},\ \bibinfo {pages} {933} (\bibinfo {year} {2012})}\BibitemShut
  {NoStop}%
\bibitem [{\citenamefont {Willke}\ \emph {et~al.}(2019)\citenamefont {Willke},
  \citenamefont {Yang}, \citenamefont {Bae}, \citenamefont {Heinrich},\ and\
  \citenamefont {Lutz}}]{Willke2019}%
  \BibitemOpen
  \bibfield  {author} {\bibinfo {author} {\bibfnamefont {P.}~\bibnamefont
  {Willke}}, \bibinfo {author} {\bibfnamefont {K.}~\bibnamefont {Yang}},
  \bibinfo {author} {\bibfnamefont {Y.}~\bibnamefont {Bae}}, \bibinfo {author}
  {\bibfnamefont {A.~J.}\ \bibnamefont {Heinrich}},\ and\ \bibinfo {author}
  {\bibfnamefont {C.~P.}\ \bibnamefont {Lutz}},\ }\bibfield  {title}
  {{\selectlanguage {en}\bibinfo {title} {Magnetic resonance imaging of single
  atoms on a surface}},\ }\href {https://doi.org/10.1038/s41567-019-0573-x}
  {\bibfield  {journal} {\bibinfo  {journal} {Nature Physics}\ }\textbf
  {\bibinfo {volume} {15}},\ \bibinfo {pages} {1005} (\bibinfo {year}
  {2019})}\BibitemShut {NoStop}%
\bibitem [{\citenamefont {Zopes}\ \emph {et~al.}(2018)\citenamefont {Zopes},
  \citenamefont {Cujia}, \citenamefont {Sasaki}, \citenamefont {Boss},
  \citenamefont {Itoh},\ and\ \citenamefont {Degen}}]{Zopes2018}%
  \BibitemOpen
  \bibfield  {author} {\bibinfo {author} {\bibfnamefont {J.}~\bibnamefont
  {Zopes}}, \bibinfo {author} {\bibfnamefont {K.~S.}\ \bibnamefont {Cujia}},
  \bibinfo {author} {\bibfnamefont {K.}~\bibnamefont {Sasaki}}, \bibinfo
  {author} {\bibfnamefont {J.~M.}\ \bibnamefont {Boss}}, \bibinfo {author}
  {\bibfnamefont {K.~M.}\ \bibnamefont {Itoh}},\ and\ \bibinfo {author}
  {\bibfnamefont {C.~L.}\ \bibnamefont {Degen}},\ }\bibfield  {title}
  {{\selectlanguage {en}\bibinfo {title} {Three-dimensional localization
  spectroscopy of individual nuclear spins with sub-{Angstrom} resolution}},\
  }\href {https://doi.org/10.1038/s41467-018-07121-0} {\bibfield  {journal}
  {\bibinfo  {journal} {Nature Communications}\ }\textbf {\bibinfo {volume}
  {9}},\ \bibinfo {pages} {4678} (\bibinfo {year} {2018})}\BibitemShut
  {NoStop}%
\bibitem [{\citenamefont {Appel}\ \emph {et~al.}(2015)\citenamefont {Appel},
  \citenamefont {Ganzhorn}, \citenamefont {Neu},\ and\ \citenamefont
  {Maletinsky}}]{Appel2015}%
  \BibitemOpen
  \bibfield  {author} {\bibinfo {author} {\bibfnamefont {P.}~\bibnamefont
  {Appel}}, \bibinfo {author} {\bibfnamefont {M.}~\bibnamefont {Ganzhorn}},
  \bibinfo {author} {\bibfnamefont {E.}~\bibnamefont {Neu}},\ and\ \bibinfo
  {author} {\bibfnamefont {P.}~\bibnamefont {Maletinsky}},\ }\bibfield  {title}
  {{\selectlanguage {en}\bibinfo {title} {Nanoscale microwave imaging with a
  single electron spin in diamond}},\ }\href
  {https://doi.org/10.1088/1367-2630/17/11/112001} {\bibfield  {journal}
  {\bibinfo  {journal} {New Journal of Physics}\ }\textbf {\bibinfo {volume}
  {17}},\ \bibinfo {pages} {112001} (\bibinfo {year} {2015})}\BibitemShut
  {NoStop}%
\bibitem [{\citenamefont {Böhi}\ \emph {et~al.}(2010)\citenamefont {Böhi},
  \citenamefont {Riedel}, \citenamefont {Hänsch},\ and\ \citenamefont
  {Treutlein}}]{Boehi2010}%
  \BibitemOpen
  \bibfield  {author} {\bibinfo {author} {\bibfnamefont {P.}~\bibnamefont
  {Böhi}}, \bibinfo {author} {\bibfnamefont {M.~F.}\ \bibnamefont {Riedel}},
  \bibinfo {author} {\bibfnamefont {T.~W.}\ \bibnamefont {Hänsch}},\ and\
  \bibinfo {author} {\bibfnamefont {P.}~\bibnamefont {Treutlein}},\ }\bibfield
  {title} {\bibinfo {title} {Imaging of microwave fields using ultracold
  atoms},\ }\href {https://doi.org/10.1063/1.3470591} {\bibfield  {journal}
  {\bibinfo  {journal} {Applied Physics Letters}\ }\textbf {\bibinfo {volume}
  {97}},\ \bibinfo {pages} {051101} (\bibinfo {year} {2010})}\BibitemShut
  {NoStop}%
\bibitem [{\citenamefont {Horsley}\ \emph {et~al.}(2015)\citenamefont
  {Horsley}, \citenamefont {Du},\ and\ \citenamefont
  {Treutlein}}]{Horsley2015}%
  \BibitemOpen
  \bibfield  {author} {\bibinfo {author} {\bibfnamefont {A.}~\bibnamefont
  {Horsley}}, \bibinfo {author} {\bibfnamefont {G.-X.}\ \bibnamefont {Du}},\
  and\ \bibinfo {author} {\bibfnamefont {P.}~\bibnamefont {Treutlein}},\
  }\bibfield  {title} {{\selectlanguage {en}\bibinfo {title} {Widefield
  microwave imaging in alkali vapor cells with sub-100 $\mu$m resolution}},\
  }\href {https://doi.org/10.1088/1367-2630/17/11/112002} {\bibfield  {journal}
  {\bibinfo  {journal} {New Journal of Physics}\ }\textbf {\bibinfo {volume}
  {17}},\ \bibinfo {pages} {112002} (\bibinfo {year} {2015})}\BibitemShut
  {NoStop}%
\bibitem [{\citenamefont {Lee}\ \emph {et~al.}(2014)\citenamefont {Lee},
  \citenamefont {Kim}, \citenamefont {Seo}, \citenamefont {Hong}, \citenamefont
  {Song}, \citenamefont {Dasari},\ and\ \citenamefont {An}}]{Lee2014}%
  \BibitemOpen
  \bibfield  {author} {\bibinfo {author} {\bibfnamefont {M.}~\bibnamefont
  {Lee}}, \bibinfo {author} {\bibfnamefont {J.}~\bibnamefont {Kim}}, \bibinfo
  {author} {\bibfnamefont {W.}~\bibnamefont {Seo}}, \bibinfo {author}
  {\bibfnamefont {H.-G.}\ \bibnamefont {Hong}}, \bibinfo {author}
  {\bibfnamefont {Y.}~\bibnamefont {Song}}, \bibinfo {author} {\bibfnamefont
  {R.~R.}\ \bibnamefont {Dasari}},\ and\ \bibinfo {author} {\bibfnamefont
  {K.}~\bibnamefont {An}},\ }\bibfield  {title} {{\selectlanguage {en}\bibinfo
  {title} {Three-dimensional imaging of cavity vacuum with single atoms
  localized by a nanohole array}},\ }\href {https://doi.org/10.1038/ncomms4441}
  {\bibfield  {journal} {\bibinfo  {journal} {Nature Communications}\ }\textbf
  {\bibinfo {volume} {5}},\ \bibinfo {pages} {3441} (\bibinfo {year}
  {2014})}\BibitemShut {NoStop}%
\bibitem [{\citenamefont {Lvovsky}\ \emph {et~al.}(2009)\citenamefont
  {Lvovsky}, \citenamefont {Sanders},\ and\ \citenamefont
  {Tittel}}]{Lvovsky2009}%
  \BibitemOpen
  \bibfield  {author} {\bibinfo {author} {\bibfnamefont {A.~I.}\ \bibnamefont
  {Lvovsky}}, \bibinfo {author} {\bibfnamefont {B.~C.}\ \bibnamefont
  {Sanders}},\ and\ \bibinfo {author} {\bibfnamefont {W.}~\bibnamefont
  {Tittel}},\ }\bibfield  {title} {{\selectlanguage {en}\bibinfo {title}
  {Optical quantum memory}},\ }\href {https://doi.org/10.1038/nphoton.2009.231}
  {\bibfield  {journal} {\bibinfo  {journal} {Nature Photonics}\ }\textbf
  {\bibinfo {volume} {3}},\ \bibinfo {pages} {706} (\bibinfo {year}
  {2009})}\BibitemShut {NoStop}%
\bibitem [{\citenamefont {Saglamyurek}\ \emph {et~al.}(2018)\citenamefont
  {Saglamyurek}, \citenamefont {Hrushevskyi}, \citenamefont {Rastogi},
  \citenamefont {Heshami},\ and\ \citenamefont {LeBlanc}}]{Saglamyurek2018}%
  \BibitemOpen
  \bibfield  {author} {\bibinfo {author} {\bibfnamefont {E.}~\bibnamefont
  {Saglamyurek}}, \bibinfo {author} {\bibfnamefont {T.}~\bibnamefont
  {Hrushevskyi}}, \bibinfo {author} {\bibfnamefont {A.}~\bibnamefont
  {Rastogi}}, \bibinfo {author} {\bibfnamefont {K.}~\bibnamefont {Heshami}},\
  and\ \bibinfo {author} {\bibfnamefont {L.~J.}\ \bibnamefont {LeBlanc}},\
  }\bibfield  {title} {{\selectlanguage {en}\bibinfo {title} {Coherent storage
  and manipulation of broadband photons via dynamically controlled
  {Autler}–{Townes} splitting}},\ }\href
  {https://doi.org/10.1038/s41566-018-0279-0} {\bibfield  {journal} {\bibinfo
  {journal} {Nature Photonics}\ }\textbf {\bibinfo {volume} {12}},\ \bibinfo
  {pages} {774} (\bibinfo {year} {2018})}\BibitemShut {NoStop}%
\bibitem [{\citenamefont {Duan}\ \emph {et~al.}(2001)\citenamefont {Duan},
  \citenamefont {Lukin}, \citenamefont {Cirac},\ and\ \citenamefont
  {Zoller}}]{Duan2001}%
  \BibitemOpen
  \bibfield  {author} {\bibinfo {author} {\bibfnamefont {L.-M.}\ \bibnamefont
  {Duan}}, \bibinfo {author} {\bibfnamefont {M.~D.}\ \bibnamefont {Lukin}},
  \bibinfo {author} {\bibfnamefont {J.~I.}\ \bibnamefont {Cirac}},\ and\
  \bibinfo {author} {\bibfnamefont {P.}~\bibnamefont {Zoller}},\ }\bibfield
  {title} {{\selectlanguage {en}\bibinfo {title} {Long-distance quantum
  communication with atomic ensembles and linear optics}},\ }\href
  {https://doi.org/10.1038/35106500} {\bibfield  {journal} {\bibinfo  {journal}
  {Nature}\ }\textbf {\bibinfo {volume} {414}},\ \bibinfo {pages} {413}
  (\bibinfo {year} {2001})}\BibitemShut {NoStop}%
\bibitem [{\citenamefont {Castellucci}\ \emph {et~al.}(2021)\citenamefont
  {Castellucci}, \citenamefont {Clark}, \citenamefont {Selyem}, \citenamefont
  {Wang},\ and\ \citenamefont {Franke-Arnold}}]{Castellucci2021}%
  \BibitemOpen
  \bibfield  {author} {\bibinfo {author} {\bibfnamefont {F.}~\bibnamefont
  {Castellucci}}, \bibinfo {author} {\bibfnamefont {T.~W.}\ \bibnamefont
  {Clark}}, \bibinfo {author} {\bibfnamefont {A.}~\bibnamefont {Selyem}},
  \bibinfo {author} {\bibfnamefont {J.}~\bibnamefont {Wang}},\ and\ \bibinfo
  {author} {\bibfnamefont {S.}~\bibnamefont {Franke-Arnold}},\ }\bibfield
  {title} {\bibinfo {title} {Atomic {Compass}: {Detecting} {3D} {Magnetic}
  {Field} {Alignment} with {Vector} {Vortex} {Light}},\ }\href
  {https://doi.org/10.1103/PhysRevLett.127.233202} {\bibfield  {journal}
  {\bibinfo  {journal} {Physical Review Letters}\ }\textbf {\bibinfo {volume}
  {127}},\ \bibinfo {pages} {233202} (\bibinfo {year} {2021})}\BibitemShut
  {NoStop}%
\bibitem [{\citenamefont {Xu}\ \emph {et~al.}(2008)\citenamefont {Xu},
  \citenamefont {Crawford}, \citenamefont {Rochester}, \citenamefont
  {Yashchuk}, \citenamefont {Budker},\ and\ \citenamefont {Pines}}]{Xu2008}%
  \BibitemOpen
  \bibfield  {author} {\bibinfo {author} {\bibfnamefont {S.}~\bibnamefont
  {Xu}}, \bibinfo {author} {\bibfnamefont {C.~W.}\ \bibnamefont {Crawford}},
  \bibinfo {author} {\bibfnamefont {S.}~\bibnamefont {Rochester}}, \bibinfo
  {author} {\bibfnamefont {V.}~\bibnamefont {Yashchuk}}, \bibinfo {author}
  {\bibfnamefont {D.}~\bibnamefont {Budker}},\ and\ \bibinfo {author}
  {\bibfnamefont {A.}~\bibnamefont {Pines}},\ }\bibfield  {title} {\bibinfo
  {title} {Submillimeter-resolution magnetic resonance imaging at the {Earth}'s
  magnetic field with an atomic magnetometer},\ }\href
  {https://doi.org/10.1103/PhysRevA.78.013404} {\bibfield  {journal} {\bibinfo
  {journal} {Physical Review A}\ }\textbf {\bibinfo {volume} {78}},\ \bibinfo
  {pages} {013404} (\bibinfo {year} {2008})}\BibitemShut {NoStop}%
\bibitem [{\citenamefont {Jing}\ \emph {et~al.}(2020)\citenamefont {Jing},
  \citenamefont {Hu}, \citenamefont {Ma}, \citenamefont {Zhang}, \citenamefont
  {Zhang}, \citenamefont {Xiao},\ and\ \citenamefont {Jia}}]{Jing2020}%
  \BibitemOpen
  \bibfield  {author} {\bibinfo {author} {\bibfnamefont {M.}~\bibnamefont
  {Jing}}, \bibinfo {author} {\bibfnamefont {Y.}~\bibnamefont {Hu}}, \bibinfo
  {author} {\bibfnamefont {J.}~\bibnamefont {Ma}}, \bibinfo {author}
  {\bibfnamefont {H.}~\bibnamefont {Zhang}}, \bibinfo {author} {\bibfnamefont
  {L.}~\bibnamefont {Zhang}}, \bibinfo {author} {\bibfnamefont
  {L.}~\bibnamefont {Xiao}},\ and\ \bibinfo {author} {\bibfnamefont
  {S.}~\bibnamefont {Jia}},\ }\bibfield  {title} {{\selectlanguage {en}\bibinfo
  {title} {Atomic superheterodyne receiver based on microwave-dressed {Rydberg}
  spectroscopy}},\ }\href {https://doi.org/10.1038/s41567-020-0918-5}
  {\bibfield  {journal} {\bibinfo  {journal} {Nature Physics}\ }\textbf
  {\bibinfo {volume} {16}},\ \bibinfo {pages} {911} (\bibinfo {year}
  {2020})}\BibitemShut {NoStop}%
\bibitem [{\citenamefont {Downes}\ \emph {et~al.}(2020)\citenamefont {Downes},
  \citenamefont {MacKellar}, \citenamefont {Whiting}, \citenamefont
  {Bourgenot}, \citenamefont {Adams},\ and\ \citenamefont
  {Weatherill}}]{PhysRevX.10.011027}%
  \BibitemOpen
  \bibfield  {author} {\bibinfo {author} {\bibfnamefont {L.~A.}\ \bibnamefont
  {Downes}}, \bibinfo {author} {\bibfnamefont {A.~R.}\ \bibnamefont
  {MacKellar}}, \bibinfo {author} {\bibfnamefont {D.~J.}\ \bibnamefont
  {Whiting}}, \bibinfo {author} {\bibfnamefont {C.}~\bibnamefont {Bourgenot}},
  \bibinfo {author} {\bibfnamefont {C.~S.}\ \bibnamefont {Adams}},\ and\
  \bibinfo {author} {\bibfnamefont {K.~J.}\ \bibnamefont {Weatherill}},\
  }\bibfield  {title} {\bibinfo {title} {Full-field terahertz imaging at
  kilohertz frame rates using atomic vapor},\ }\href
  {https://doi.org/10.1103/PhysRevX.10.011027} {\bibfield  {journal} {\bibinfo
  {journal} {Phys. Rev. X}\ }\textbf {\bibinfo {volume} {10}},\ \bibinfo
  {pages} {011027} (\bibinfo {year} {2020})}\BibitemShut {NoStop}%
\bibitem [{\citenamefont {Sompet}\ \emph {et~al.}(2022)\citenamefont {Sompet},
  \citenamefont {Hirthe}, \citenamefont {Bourgund}, \citenamefont {Chalopin},
  \citenamefont {Bibo}, \citenamefont {Koepsell}, \citenamefont {Bojović},
  \citenamefont {Verresen}, \citenamefont {Pollmann}, \citenamefont {Salomon},
  \citenamefont {Gross}, \citenamefont {Hilker},\ and\ \citenamefont
  {Bloch}}]{Sompet2022}%
  \BibitemOpen
  \bibfield  {author} {\bibinfo {author} {\bibfnamefont {P.}~\bibnamefont
  {Sompet}}, \bibinfo {author} {\bibfnamefont {S.}~\bibnamefont {Hirthe}},
  \bibinfo {author} {\bibfnamefont {D.}~\bibnamefont {Bourgund}}, \bibinfo
  {author} {\bibfnamefont {T.}~\bibnamefont {Chalopin}}, \bibinfo {author}
  {\bibfnamefont {J.}~\bibnamefont {Bibo}}, \bibinfo {author} {\bibfnamefont
  {J.}~\bibnamefont {Koepsell}}, \bibinfo {author} {\bibfnamefont
  {P.}~\bibnamefont {Bojović}}, \bibinfo {author} {\bibfnamefont
  {R.}~\bibnamefont {Verresen}}, \bibinfo {author} {\bibfnamefont
  {F.}~\bibnamefont {Pollmann}}, \bibinfo {author} {\bibfnamefont
  {G.}~\bibnamefont {Salomon}}, \bibinfo {author} {\bibfnamefont
  {C.}~\bibnamefont {Gross}}, \bibinfo {author} {\bibfnamefont {T.~A.}\
  \bibnamefont {Hilker}},\ and\ \bibinfo {author} {\bibfnamefont
  {I.}~\bibnamefont {Bloch}},\ }\bibfield  {title} {{\selectlanguage
  {en}\bibinfo {title} {Realizing the symmetry-protected {Haldane} phase in
  {Fermi}–{Hubbard} ladders}},\ }\href
  {https://doi.org/10.1038/s41586-022-04688-z} {\bibfield  {journal} {\bibinfo
  {journal} {Nature}\ }\textbf {\bibinfo {volume} {606}},\ \bibinfo {pages}
  {484} (\bibinfo {year} {2022})}\BibitemShut {NoStop}%
\bibitem [{\citenamefont {Ferreira-Cao}\ \emph {et~al.}(2020)\citenamefont
  {Ferreira-Cao}, \citenamefont {Gavryusev}, \citenamefont {Franz},
  \citenamefont {Alves}, \citenamefont {Signoles}, \citenamefont {Zürn},\ and\
  \citenamefont {Weidemüller}}]{Ferreira-Cao_2020}%
  \BibitemOpen
  \bibfield  {author} {\bibinfo {author} {\bibfnamefont {M.}~\bibnamefont
  {Ferreira-Cao}}, \bibinfo {author} {\bibfnamefont {V.}~\bibnamefont
  {Gavryusev}}, \bibinfo {author} {\bibfnamefont {T.}~\bibnamefont {Franz}},
  \bibinfo {author} {\bibfnamefont {R.~F.}\ \bibnamefont {Alves}}, \bibinfo
  {author} {\bibfnamefont {A.}~\bibnamefont {Signoles}}, \bibinfo {author}
  {\bibfnamefont {G.}~\bibnamefont {Zürn}},\ and\ \bibinfo {author}
  {\bibfnamefont {M.}~\bibnamefont {Weidemüller}},\ }\bibfield  {title}
  {\bibinfo {title} {Depletion imaging of rydberg atoms in cold atomic gases},\
  }\href {https://doi.org/10.1088/1361-6455/ab7427} {\bibfield  {journal}
  {\bibinfo  {journal} {Journal of Physics B: Atomic, Molecular and Optical
  Physics}\ }\textbf {\bibinfo {volume} {53}},\ \bibinfo {pages} {084004}
  (\bibinfo {year} {2020})}\BibitemShut {NoStop}%
\bibitem [{\citenamefont {Bakr}\ \emph {et~al.}(2009)\citenamefont {Bakr},
  \citenamefont {Gillen}, \citenamefont {Peng}, \citenamefont {Fölling},\ and\
  \citenamefont {Greiner}}]{Bakr2009}%
  \BibitemOpen
  \bibfield  {author} {\bibinfo {author} {\bibfnamefont {W.~S.}\ \bibnamefont
  {Bakr}}, \bibinfo {author} {\bibfnamefont {J.~I.}\ \bibnamefont {Gillen}},
  \bibinfo {author} {\bibfnamefont {A.}~\bibnamefont {Peng}}, \bibinfo {author}
  {\bibfnamefont {S.}~\bibnamefont {Fölling}},\ and\ \bibinfo {author}
  {\bibfnamefont {M.}~\bibnamefont {Greiner}},\ }\bibfield  {title}
  {{\selectlanguage {en}\bibinfo {title} {A quantum gas microscope for
  detecting single atoms in a {Hubbard}-regime optical lattice}},\ }\href
  {https://doi.org/10.1038/nature08482} {\bibfield  {journal} {\bibinfo
  {journal} {Nature}\ }\textbf {\bibinfo {volume} {462}},\ \bibinfo {pages}
  {74} (\bibinfo {year} {2009})}\BibitemShut {NoStop}%
\bibitem [{\citenamefont {Cheuk}\ \emph {et~al.}(2015)\citenamefont {Cheuk},
  \citenamefont {Nichols}, \citenamefont {Okan}, \citenamefont {Gersdorf},
  \citenamefont {Ramasesh}, \citenamefont {Bakr}, \citenamefont {Lompe},\ and\
  \citenamefont {Zwierlein}}]{Cheuk2015}%
  \BibitemOpen
  \bibfield  {author} {\bibinfo {author} {\bibfnamefont {L.~W.}\ \bibnamefont
  {Cheuk}}, \bibinfo {author} {\bibfnamefont {M.~A.}\ \bibnamefont {Nichols}},
  \bibinfo {author} {\bibfnamefont {M.}~\bibnamefont {Okan}}, \bibinfo {author}
  {\bibfnamefont {T.}~\bibnamefont {Gersdorf}}, \bibinfo {author}
  {\bibfnamefont {V.~V.}\ \bibnamefont {Ramasesh}}, \bibinfo {author}
  {\bibfnamefont {W.~S.}\ \bibnamefont {Bakr}}, \bibinfo {author}
  {\bibfnamefont {T.}~\bibnamefont {Lompe}},\ and\ \bibinfo {author}
  {\bibfnamefont {M.~W.}\ \bibnamefont {Zwierlein}},\ }\bibfield  {title}
  {\bibinfo {title} {Quantum-{Gas} {Microscope} for {Fermionic} {Atoms}},\
  }\href {https://doi.org/10.1103/PhysRevLett.114.193001} {\bibfield  {journal}
  {\bibinfo  {journal} {Physical Review Letters}\ }\textbf {\bibinfo {volume}
  {114}},\ \bibinfo {pages} {193001} (\bibinfo {year} {2015})}\BibitemShut
  {NoStop}%
\bibitem [{\citenamefont {Balasubramanian}\ \emph {et~al.}(2008)\citenamefont
  {Balasubramanian}, \citenamefont {Chan}, \citenamefont {Kolesov},
  \citenamefont {Al-Hmoud}, \citenamefont {Tisler}, \citenamefont {Shin},
  \citenamefont {Kim}, \citenamefont {Wojcik}, \citenamefont {Hemmer},
  \citenamefont {Krueger}, \citenamefont {Hanke}, \citenamefont
  {Leitenstorfer}, \citenamefont {Bratschitsch}, \citenamefont {Jelezko},\ and\
  \citenamefont {Wrachtrup}}]{Balasubramanian2008}%
  \BibitemOpen
  \bibfield  {author} {\bibinfo {author} {\bibfnamefont {G.}~\bibnamefont
  {Balasubramanian}}, \bibinfo {author} {\bibfnamefont {I.~Y.}\ \bibnamefont
  {Chan}}, \bibinfo {author} {\bibfnamefont {R.}~\bibnamefont {Kolesov}},
  \bibinfo {author} {\bibfnamefont {M.}~\bibnamefont {Al-Hmoud}}, \bibinfo
  {author} {\bibfnamefont {J.}~\bibnamefont {Tisler}}, \bibinfo {author}
  {\bibfnamefont {C.}~\bibnamefont {Shin}}, \bibinfo {author} {\bibfnamefont
  {C.}~\bibnamefont {Kim}}, \bibinfo {author} {\bibfnamefont {A.}~\bibnamefont
  {Wojcik}}, \bibinfo {author} {\bibfnamefont {P.~R.}\ \bibnamefont {Hemmer}},
  \bibinfo {author} {\bibfnamefont {A.}~\bibnamefont {Krueger}}, \bibinfo
  {author} {\bibfnamefont {T.}~\bibnamefont {Hanke}}, \bibinfo {author}
  {\bibfnamefont {A.}~\bibnamefont {Leitenstorfer}}, \bibinfo {author}
  {\bibfnamefont {R.}~\bibnamefont {Bratschitsch}}, \bibinfo {author}
  {\bibfnamefont {F.}~\bibnamefont {Jelezko}},\ and\ \bibinfo {author}
  {\bibfnamefont {J.}~\bibnamefont {Wrachtrup}},\ }\bibfield  {title}
  {{\selectlanguage {en}\bibinfo {title} {Nanoscale imaging magnetometry with
  diamond spins under ambient conditions}},\ }\href
  {https://doi.org/10.1038/nature07278} {\bibfield  {journal} {\bibinfo
  {journal} {Nature}\ }\textbf {\bibinfo {volume} {455}},\ \bibinfo {pages}
  {648} (\bibinfo {year} {2008})}\BibitemShut {NoStop}%
\bibitem [{\citenamefont {Gruber}\ \emph {et~al.}(1997)\citenamefont {Gruber},
  \citenamefont {Dräbenstedt}, \citenamefont {Tietz}, \citenamefont {Fleury},
  \citenamefont {Wrachtrup},\ and\ \citenamefont {Borczyskowski}}]{Gruber1997}%
  \BibitemOpen
  \bibfield  {author} {\bibinfo {author} {\bibfnamefont {A.}~\bibnamefont
  {Gruber}}, \bibinfo {author} {\bibfnamefont {A.}~\bibnamefont
  {Dräbenstedt}}, \bibinfo {author} {\bibfnamefont {C.}~\bibnamefont {Tietz}},
  \bibinfo {author} {\bibfnamefont {L.}~\bibnamefont {Fleury}}, \bibinfo
  {author} {\bibfnamefont {J.}~\bibnamefont {Wrachtrup}},\ and\ \bibinfo
  {author} {\bibfnamefont {C.~v.}\ \bibnamefont {Borczyskowski}},\ }\bibfield
  {title} {\bibinfo {title} {Scanning {Confocal} {Optical} {Microscopy} and
  {Magnetic} {Resonance} on {Single} {Defect} {Centers}},\ }\href
  {https://doi.org/10.1126/science.276.5321.2012} {\bibfield  {journal}
  {\bibinfo  {journal} {Science}\ }\textbf {\bibinfo {volume} {276}},\ \bibinfo
  {pages} {2012} (\bibinfo {year} {1997})}\BibitemShut {NoStop}%
\bibitem [{\citenamefont {Hsiao}\ \emph {et~al.}(2018)\citenamefont {Hsiao},
  \citenamefont {Tsai}, \citenamefont {Chen}, \citenamefont {Lin},
  \citenamefont {Hung}, \citenamefont {Lee}, \citenamefont {Chen},
  \citenamefont {Chen}, \citenamefont {Yu},\ and\ \citenamefont
  {Chen}}]{Hsiao2018}%
  \BibitemOpen
  \bibfield  {author} {\bibinfo {author} {\bibfnamefont {Y.-F.}\ \bibnamefont
  {Hsiao}}, \bibinfo {author} {\bibfnamefont {P.-J.}\ \bibnamefont {Tsai}},
  \bibinfo {author} {\bibfnamefont {H.-S.}\ \bibnamefont {Chen}}, \bibinfo
  {author} {\bibfnamefont {S.-X.}\ \bibnamefont {Lin}}, \bibinfo {author}
  {\bibfnamefont {C.-C.}\ \bibnamefont {Hung}}, \bibinfo {author}
  {\bibfnamefont {C.-H.}\ \bibnamefont {Lee}}, \bibinfo {author} {\bibfnamefont
  {Y.-H.}\ \bibnamefont {Chen}}, \bibinfo {author} {\bibfnamefont {Y.-F.}\
  \bibnamefont {Chen}}, \bibinfo {author} {\bibfnamefont {I.~A.}\ \bibnamefont
  {Yu}},\ and\ \bibinfo {author} {\bibfnamefont {Y.-C.}\ \bibnamefont {Chen}},\
  }\bibfield  {title} {\bibinfo {title} {Highly {Efficient} {Coherent}
  {Optical} {Memory} {Based} on {Electromagnetically} {Induced}
  {Transparency}},\ }\href {https://doi.org/10.1103/PhysRevLett.120.183602}
  {\bibfield  {journal} {\bibinfo  {journal} {Physical Review Letters}\
  }\textbf {\bibinfo {volume} {120}},\ \bibinfo {pages} {183602} (\bibinfo
  {year} {2018})}\BibitemShut {NoStop}%
\bibitem [{\citenamefont {Guo}\ \emph {et~al.}(2019)\citenamefont {Guo},
  \citenamefont {Feng}, \citenamefont {Yang}, \citenamefont {Yu}, \citenamefont
  {Chen}, \citenamefont {Yuan},\ and\ \citenamefont {Zhang}}]{Guo2019}%
  \BibitemOpen
  \bibfield  {author} {\bibinfo {author} {\bibfnamefont {J.}~\bibnamefont
  {Guo}}, \bibinfo {author} {\bibfnamefont {X.}~\bibnamefont {Feng}}, \bibinfo
  {author} {\bibfnamefont {P.}~\bibnamefont {Yang}}, \bibinfo {author}
  {\bibfnamefont {Z.}~\bibnamefont {Yu}}, \bibinfo {author} {\bibfnamefont
  {L.~Q.}\ \bibnamefont {Chen}}, \bibinfo {author} {\bibfnamefont {C.-H.}\
  \bibnamefont {Yuan}},\ and\ \bibinfo {author} {\bibfnamefont
  {W.}~\bibnamefont {Zhang}},\ }\bibfield  {title} {{\selectlanguage
  {en}\bibinfo {title} {High-performance {Raman} quantum memory with optimal
  control in room temperature atoms}},\ }\href
  {https://doi.org/10.1038/s41467-018-08118-5} {\bibfield  {journal} {\bibinfo
  {journal} {Nature Communications}\ }\textbf {\bibinfo {volume} {10}},\
  \bibinfo {pages} {148} (\bibinfo {year} {2019})}\BibitemShut {NoStop}%
\bibitem [{\citenamefont {Cho}\ \emph {et~al.}(2016)\citenamefont {Cho},
  \citenamefont {Campbell}, \citenamefont {Everett}, \citenamefont {Bernu},
  \citenamefont {Higginbottom}, \citenamefont {Cao}, \citenamefont {Geng},
  \citenamefont {Robins}, \citenamefont {Lam},\ and\ \citenamefont
  {Buchler}}]{Cho2016}%
  \BibitemOpen
  \bibfield  {author} {\bibinfo {author} {\bibfnamefont {Y.-W.}\ \bibnamefont
  {Cho}}, \bibinfo {author} {\bibfnamefont {G.~T.}\ \bibnamefont {Campbell}},
  \bibinfo {author} {\bibfnamefont {J.~L.}\ \bibnamefont {Everett}}, \bibinfo
  {author} {\bibfnamefont {J.}~\bibnamefont {Bernu}}, \bibinfo {author}
  {\bibfnamefont {D.~B.}\ \bibnamefont {Higginbottom}}, \bibinfo {author}
  {\bibfnamefont {M.~T.}\ \bibnamefont {Cao}}, \bibinfo {author} {\bibfnamefont
  {J.}~\bibnamefont {Geng}}, \bibinfo {author} {\bibfnamefont {N.~P.}\
  \bibnamefont {Robins}}, \bibinfo {author} {\bibfnamefont {P.~K.}\
  \bibnamefont {Lam}},\ and\ \bibinfo {author} {\bibfnamefont {B.~C.}\
  \bibnamefont {Buchler}},\ }\bibfield  {title} {{\selectlanguage {EN}\bibinfo
  {title} {Highly efficient optical quantum memory with long coherence time in
  cold atoms}},\ }\href {https://doi.org/10.1364/OPTICA.3.000100} {\bibfield
  {journal} {\bibinfo  {journal} {Optica}\ }\textbf {\bibinfo {volume} {3}},\
  \bibinfo {pages} {100} (\bibinfo {year} {2016})}\BibitemShut {NoStop}%
\bibitem [{\citenamefont {Lipka}\ \emph {et~al.}(2019)\citenamefont {Lipka},
  \citenamefont {Leszczyński}, \citenamefont {Mazelanik}, \citenamefont
  {Parniak},\ and\ \citenamefont {Wasilewski}}]{Lipka2019}%
  \BibitemOpen
  \bibfield  {author} {\bibinfo {author} {\bibfnamefont {M.}~\bibnamefont
  {Lipka}}, \bibinfo {author} {\bibfnamefont {A.}~\bibnamefont {Leszczyński}},
  \bibinfo {author} {\bibfnamefont {M.}~\bibnamefont {Mazelanik}}, \bibinfo
  {author} {\bibfnamefont {M.}~\bibnamefont {Parniak}},\ and\ \bibinfo {author}
  {\bibfnamefont {W.}~\bibnamefont {Wasilewski}},\ }\bibfield  {title}
  {\bibinfo {title} {Spatial {Spin}-{Wave} {Modulator} for
  {Quantum}-{Memory}-{Assisted} {Adaptive} {Measurements}},\ }\href
  {https://doi.org/10.1103/PhysRevApplied.11.034049} {\bibfield  {journal}
  {\bibinfo  {journal} {Physical Review Applied}\ }\textbf {\bibinfo {volume}
  {11}},\ \bibinfo {pages} {034049} (\bibinfo {year} {2019})}\BibitemShut
  {NoStop}%
\bibitem [{\citenamefont {Mazelanik}\ \emph {et~al.}(2019)\citenamefont
  {Mazelanik}, \citenamefont {Parniak}, \citenamefont {Leszczyński},
  \citenamefont {Lipka},\ and\ \citenamefont {Wasilewski}}]{Mazelanik2019}%
  \BibitemOpen
  \bibfield  {author} {\bibinfo {author} {\bibfnamefont {M.}~\bibnamefont
  {Mazelanik}}, \bibinfo {author} {\bibfnamefont {M.}~\bibnamefont {Parniak}},
  \bibinfo {author} {\bibfnamefont {A.}~\bibnamefont {Leszczyński}}, \bibinfo
  {author} {\bibfnamefont {M.}~\bibnamefont {Lipka}},\ and\ \bibinfo {author}
  {\bibfnamefont {W.}~\bibnamefont {Wasilewski}},\ }\bibfield  {title}
  {{\selectlanguage {en}\bibinfo {title} {Coherent spin-wave processor of
  stored optical pulses}},\ }\href@noop {} {\bibfield  {journal} {\bibinfo
  {journal} {npj Quantum Information}\ }\textbf {\bibinfo {volume} {5}},\
  \bibinfo {pages} {22} (\bibinfo {year} {2019})}\BibitemShut {NoStop}%
\bibitem [{\citenamefont {Parniak}\ \emph {et~al.}(2019)\citenamefont
  {Parniak}, \citenamefont {Mazelanik}, \citenamefont {Leszczyński},
  \citenamefont {Lipka}, \citenamefont {Dąbrowski},\ and\ \citenamefont
  {Wasilewski}}]{Parniak2019}%
  \BibitemOpen
  \bibfield  {author} {\bibinfo {author} {\bibfnamefont {M.}~\bibnamefont
  {Parniak}}, \bibinfo {author} {\bibfnamefont {M.}~\bibnamefont {Mazelanik}},
  \bibinfo {author} {\bibfnamefont {A.}~\bibnamefont {Leszczyński}}, \bibinfo
  {author} {\bibfnamefont {M.}~\bibnamefont {Lipka}}, \bibinfo {author}
  {\bibfnamefont {M.}~\bibnamefont {Dąbrowski}},\ and\ \bibinfo {author}
  {\bibfnamefont {W.}~\bibnamefont {Wasilewski}},\ }\bibfield  {title}
  {\bibinfo {title} {Quantum {Optics} of {Spin} {Waves} through ac {Stark}
  {Modulation}},\ }\href {https://doi.org/10.1103/PhysRevLett.122.063604}
  {\bibfield  {journal} {\bibinfo  {journal} {Physical Review Letters}\
  }\textbf {\bibinfo {volume} {122}},\ \bibinfo {pages} {063604} (\bibinfo
  {year} {2019})}\BibitemShut {NoStop}%
\bibitem [{\citenamefont {Mazelanik}\ \emph
  {et~al.}(2023{\natexlab{a}})\citenamefont {Mazelanik}, \citenamefont
  {Leszczyński}, \citenamefont {Szawełło},\ and\ \citenamefont
  {Parniak}}]{Mazelanik2023}%
  \BibitemOpen
  \bibfield  {author} {\bibinfo {author} {\bibfnamefont {M.}~\bibnamefont
  {Mazelanik}}, \bibinfo {author} {\bibfnamefont {A.}~\bibnamefont
  {Leszczyński}}, \bibinfo {author} {\bibfnamefont {T.}~\bibnamefont
  {Szawełło}},\ and\ \bibinfo {author} {\bibfnamefont {M.}~\bibnamefont
  {Parniak}},\ }\href {https://doi.org/10.6084/m9.figshare.21804069} {\bibinfo
  {title} {Supplementary media for: Coherent optical two-photon resonance
  tomographic imaging in three dimensions,
  \url{https://dx.doi.org/10.6084/m9.figshare.21804069}}} (\bibinfo {year}
  {2023}{\natexlab{a}})\BibitemShut {NoStop}%
\bibitem [{\citenamefont {Mazelanik}\ \emph {et~al.}(2020)\citenamefont
  {Mazelanik}, \citenamefont {Leszczyński}, \citenamefont {Lipka},
  \citenamefont {Parniak},\ and\ \citenamefont {Wasilewski}}]{Mazelanik2020}%
  \BibitemOpen
  \bibfield  {author} {\bibinfo {author} {\bibfnamefont {M.}~\bibnamefont
  {Mazelanik}}, \bibinfo {author} {\bibfnamefont {A.}~\bibnamefont
  {Leszczyński}}, \bibinfo {author} {\bibfnamefont {M.}~\bibnamefont {Lipka}},
  \bibinfo {author} {\bibfnamefont {M.}~\bibnamefont {Parniak}},\ and\ \bibinfo
  {author} {\bibfnamefont {W.}~\bibnamefont {Wasilewski}},\ }\bibfield  {title}
  {{\selectlanguage {EN}\bibinfo {title} {Temporal imaging for ultra-narrowband
  few-photon states of light}},\ }\href {https://doi.org/10.1364/OPTICA.382891}
  {\bibfield  {journal} {\bibinfo  {journal} {Optica}\ }\textbf {\bibinfo
  {volume} {7}},\ \bibinfo {pages} {203} (\bibinfo {year} {2020})}\BibitemShut
  {NoStop}%
\bibitem [{\citenamefont {Mazelanik}\ \emph {et~al.}(2022)\citenamefont
  {Mazelanik}, \citenamefont {Leszczyński},\ and\ \citenamefont
  {Parniak}}]{Mazelanik2022}%
  \BibitemOpen
  \bibfield  {author} {\bibinfo {author} {\bibfnamefont {M.}~\bibnamefont
  {Mazelanik}}, \bibinfo {author} {\bibfnamefont {A.}~\bibnamefont
  {Leszczyński}},\ and\ \bibinfo {author} {\bibfnamefont {M.}~\bibnamefont
  {Parniak}},\ }\bibfield  {title} {{\selectlanguage {en}\bibinfo {title}
  {Optical-domain spectral super-resolution via a quantum-memory-based
  time-frequency processor}},\ }\href
  {https://doi.org/10.1038/s41467-022-28066-5} {\bibfield  {journal} {\bibinfo
  {journal} {Nature Communications}\ }\textbf {\bibinfo {volume} {13}},\
  \bibinfo {pages} {691} (\bibinfo {year} {2022})}\BibitemShut {NoStop}%
\bibitem [{\citenamefont {Parniak}\ \emph {et~al.}(2017)\citenamefont
  {Parniak}, \citenamefont {Dąbrowski}, \citenamefont {Mazelanik},
  \citenamefont {Leszczyński}, \citenamefont {Lipka},\ and\ \citenamefont
  {Wasilewski}}]{Parniak2017}%
  \BibitemOpen
  \bibfield  {author} {\bibinfo {author} {\bibfnamefont {M.}~\bibnamefont
  {Parniak}}, \bibinfo {author} {\bibfnamefont {M.}~\bibnamefont {Dąbrowski}},
  \bibinfo {author} {\bibfnamefont {M.}~\bibnamefont {Mazelanik}}, \bibinfo
  {author} {\bibfnamefont {A.}~\bibnamefont {Leszczyński}}, \bibinfo {author}
  {\bibfnamefont {M.}~\bibnamefont {Lipka}},\ and\ \bibinfo {author}
  {\bibfnamefont {W.}~\bibnamefont {Wasilewski}},\ }\bibfield  {title}
  {{\selectlanguage {en}\bibinfo {title} {Wavevector multiplexed atomic quantum
  memory via spatially-resolved single-photon detection}},\ }\href
  {https://doi.org/10.1038/s41467-017-02366-7} {\bibfield  {journal} {\bibinfo
  {journal} {Nature Communications}\ }\textbf {\bibinfo {volume} {8}},\
  \bibinfo {pages} {2140} (\bibinfo {year} {2017})}\BibitemShut {NoStop}%
\bibitem [{\citenamefont {Lipka}\ \emph {et~al.}(2021)\citenamefont {Lipka},
  \citenamefont {Mazelanik}, \citenamefont {Leszczyński}, \citenamefont
  {Wasilewski},\ and\ \citenamefont {Parniak}}]{Lipka2021}%
  \BibitemOpen
  \bibfield  {author} {\bibinfo {author} {\bibfnamefont {M.}~\bibnamefont
  {Lipka}}, \bibinfo {author} {\bibfnamefont {M.}~\bibnamefont {Mazelanik}},
  \bibinfo {author} {\bibfnamefont {A.}~\bibnamefont {Leszczyński}}, \bibinfo
  {author} {\bibfnamefont {W.}~\bibnamefont {Wasilewski}},\ and\ \bibinfo
  {author} {\bibfnamefont {M.}~\bibnamefont {Parniak}},\ }\bibfield  {title}
  {{\selectlanguage {en}\bibinfo {title} {Massively-multiplexed generation of
  {Bell}-type entanglement using a quantum memory}},\ }\href
  {https://doi.org/10.1038/s42005-021-00551-1} {\bibfield  {journal} {\bibinfo
  {journal} {Communications Physics}\ }\textbf {\bibinfo {volume} {4}},\
  \bibinfo {pages} {46} (\bibinfo {year} {2021})}\BibitemShut {NoStop}%
\bibitem [{\citenamefont {Strauss}(1994)}]{Strauss1994}%
  \BibitemOpen
  \bibfield  {author} {\bibinfo {author} {\bibfnamefont {C.~E.~M.}\
  \bibnamefont {Strauss}},\ }\bibfield  {title} {\bibinfo {title}
  {Synthetic-array heterodyne detection: a single-element detector acts as an
  array},\ }\href {https://doi.org/10.1364/OL.19.001609} {\bibfield  {journal}
  {\bibinfo  {journal} {Opt. Lett.}\ }\textbf {\bibinfo {volume} {19}},\
  \bibinfo {pages} {1609} (\bibinfo {year} {1994})}\BibitemShut {NoStop}%
\bibitem [{\citenamefont {Sedlacek}\ \emph {et~al.}(2012)\citenamefont
  {Sedlacek}, \citenamefont {Schwettmann}, \citenamefont {K{\"u}bler},
  \citenamefont {L{\"o}w}, \citenamefont {Pfau},\ and\ \citenamefont
  {Shaffer}}]{Sedlacek2012}%
  \BibitemOpen
  \bibfield  {author} {\bibinfo {author} {\bibfnamefont {J.~A.}\ \bibnamefont
  {Sedlacek}}, \bibinfo {author} {\bibfnamefont {A.}~\bibnamefont
  {Schwettmann}}, \bibinfo {author} {\bibfnamefont {H.}~\bibnamefont
  {K{\"u}bler}}, \bibinfo {author} {\bibfnamefont {R.}~\bibnamefont {L{\"o}w}},
  \bibinfo {author} {\bibfnamefont {T.}~\bibnamefont {Pfau}},\ and\ \bibinfo
  {author} {\bibfnamefont {J.~P.}\ \bibnamefont {Shaffer}},\ }\bibfield
  {title} {\bibinfo {title} {Microwave electrometry with rydberg atoms in a
  vapour cell using bright atomic resonances},\ }\href
  {https://doi.org/10.1038/nphys2423} {\bibfield  {journal} {\bibinfo
  {journal} {Nature Physics}\ }\textbf {\bibinfo {volume} {8}},\ \bibinfo
  {pages} {819} (\bibinfo {year} {2012})}\BibitemShut {NoStop}%
\bibitem [{\citenamefont {Gullans}\ \emph {et~al.}(2017)\citenamefont
  {Gullans}, \citenamefont {Diehl}, \citenamefont {Rittenhouse}, \citenamefont
  {Ruzic}, \citenamefont {D'Incao}, \citenamefont {Julienne}, \citenamefont
  {Gorshkov},\ and\ \citenamefont {Taylor}}]{Gullans2017}%
  \BibitemOpen
  \bibfield  {author} {\bibinfo {author} {\bibfnamefont {M.~J.}\ \bibnamefont
  {Gullans}}, \bibinfo {author} {\bibfnamefont {S.}~\bibnamefont {Diehl}},
  \bibinfo {author} {\bibfnamefont {S.~T.}\ \bibnamefont {Rittenhouse}},
  \bibinfo {author} {\bibfnamefont {B.~P.}\ \bibnamefont {Ruzic}}, \bibinfo
  {author} {\bibfnamefont {J.~P.}\ \bibnamefont {D'Incao}}, \bibinfo {author}
  {\bibfnamefont {P.}~\bibnamefont {Julienne}}, \bibinfo {author}
  {\bibfnamefont {A.~V.}\ \bibnamefont {Gorshkov}},\ and\ \bibinfo {author}
  {\bibfnamefont {J.~M.}\ \bibnamefont {Taylor}},\ }\bibfield  {title}
  {\bibinfo {title} {Efimov states of strongly interacting photons},\ }\href
  {https://doi.org/10.1103/PhysRevLett.119.233601} {\bibfield  {journal}
  {\bibinfo  {journal} {Phys. Rev. Lett.}\ }\textbf {\bibinfo {volume} {119}},\
  \bibinfo {pages} {233601} (\bibinfo {year} {2017})}\BibitemShut {NoStop}%
\bibitem [{\citenamefont {Mazelanik}\ \emph
  {et~al.}(2023{\natexlab{b}})\citenamefont {Mazelanik}, \citenamefont
  {Leszczyński}, \citenamefont {Szawełło},\ and\ \citenamefont
  {Parniak}}]{dane}%
  \BibitemOpen
  \bibfield  {author} {\bibinfo {author} {\bibfnamefont {M.}~\bibnamefont
  {Mazelanik}}, \bibinfo {author} {\bibfnamefont {A.}~\bibnamefont
  {Leszczyński}}, \bibinfo {author} {\bibfnamefont {T.}~\bibnamefont
  {Szawełło}},\ and\ \bibinfo {author} {\bibfnamefont {M.}~\bibnamefont
  {Parniak}},\ }\href {https://doi.org/10.7910/DVN/HOMGFB} {\bibinfo {title}
  {{Data for: Coherent optical two-photon resonance tomographic imaging in
  three dimensions, \url{https://doi.org/10.7910/DVN/HOMGFB}}}} (\bibinfo
  {year} {2023}{\natexlab{b}})\BibitemShut {NoStop}%
\end{thebibliography}%

\end{document}